\documentclass{article}
\usepackage[utf8]{inputenc}
\usepackage[T1]{fontenc}
\usepackage[toc,page]{appendix}
\usepackage[hang]{footmisc}
\usepackage{graphicx, fancyhdr, amssymb, amsmath, geometry, lscape, theorem, xcolor, float, rotating, setspace, authblk, longtable, lineno, ulem, enumitem,booktabs}
\usepackage{bbm}
\usepackage{comment}
\usepackage{subcaption}
\usepackage{array}
\usepackage{rotfloat}
\usepackage{units}
\usepackage{textcomp}
\usepackage{mathtools}
\usepackage{graphicx}
\usepackage{microtype}
\usepackage{graphicx}
\usepackage{caption}
\usepackage{hyperref}

\newcommand{\bi}{\begin{itemize}}
\newcommand{\ei}{\end{itemize}}
\newcommand{\beq}{\begin{equation}}
\newcommand{\eeq}{\end{equation}}
\newcommand{\be}{\begin{equation}}
\newcommand{\ee}{\end{equation}}

\definecolor{winered}{rgb}{0.5,0,0}

\begin{document}
\onehalfspacing

\title{Satellite and Mobile Phone Data Reveal How Violence Affects Seasonal Migration in Afghanistan}

\author[1]{Xiao Hui Tai}
\author[2]{Suraj R. Nair}
\author[2]{Shikhar Mehra}
\author[2,*]{Joshua E. Blumenstock}

\affil[1]{Department of Statistics, University of California, Davis, CA 95616}
\affil[2]{School of Information, University of California, Berkeley, CA 94720}
\affil[*]{Corresponding author: jblumenstock@berkeley.edu}
% \affil[c]{Affiliation Three}

\date{This version of text: \\ \today}
%\keywords{internal displacement, violence, conflict, mobile phone data} 
\maketitle

\begin{abstract}
\noindent 
Seasonal migration plays a critical role in stabilizing rural economies and sustaining the livelihoods of agricultural households. Violence and civil conflict have long been thought to disrupt these labor flows, but this hypothesis has historically been hard to test given the lack of reliable data on migration in conflict zones. Focusing on Afghanistan in the 8-year period prior to the Taliban’s takeover in 2021, we first demonstrate how satellite imagery can be used to infer the timing of the opium harvest, which employs a large number of seasonal workers in relatively well-paid jobs. We then use a dataset of nationwide mobile phone records to characterize the migration response to this harvest, and examine whether and how violence and civil conflict disrupt this migration. We find that, on average, districts with high levels of poppy cultivation receive significantly more seasonal migrants than districts with no poppy cultivation. These labor flows are surprisingly resilient to idiosyncratic violent events at the source or destination, including extreme violence resulting in large numbers of fatalities. However, seasonal migration is affected by longer-term patterns of conflict, such as the extent of Taliban control in origin and destination locations.
\end{abstract}

\vspace{5mm}


%\linenumbers

\section*{Introduction} 
\label{sec:intro}

Migrants play a central role in helping an economy make efficient use of its resources \cite{harris1970j,borjas1999economic}.  Migrant labor can increase the productivity of farms and agriculture; the income from seasonal migration also helps sustain the livelihoods of hundreds of millions of households globally \cite{bryan2014underinvestment, rakotonarivo2021ilo}. 

Civil conflict, which impacts many aspects of society and the economy, likewise threatens to disrupt seasonal labor flows. With roughly 1 billion people living in fragile and conflict-affected states \cite{world2023world}, understanding the relationship between conflict and migration has important implications for the design and targeting of public policies that ameliorate the impacts of conflict. This relationship is also theoretically ambiguous: on the one hand, conflict can increase migration from affected regions \cite{engel2007displacement,adhikari_conflict-induced_2013, tai2022mobile}; on the other, it may deter migration if people avoid the risks of travel or if mobility is restricted by checkpoints, roadblocks and other barriers \cite{melnikov2020gangs,ksoll2021guns,cali2018labor,mansour2010effects}.

Characterizing the impact of conflict on labor migration has historically been difficult in part because of the challenges inherent to data collection in insecure regions, where traditional statistical infrastructure is frequently limited and it is often difficult to conduct new surveys. These challenges are compounded by the inherent difficulty of tracking migrants who, by definition, are on the move \cite{bell2015internal,chi2020general}. 

Here we use multiple sources of non-traditional data to characterize the role of conflict in seasonal migration in Afghanistan. Our first contribution is methodological, and shows how seasonal labor flows can be inferred from satellite imagery and mobile phone data. We use high-resolution daytime satellite imagery to determine the timing of peak agricultural activity in each district; for the main crops in Afghanistan, this occurs roughly two weeks before harvest. We then use the mobile phone data to identify the seasonal migrants who migrate into districts temporarily during the harvest seasons. This approach builds on recent work to measure agricultural output, including opium poppy, from satellite imagery \cite{simms2014application,wang2019crop,koirala2024mapping, mansfield_will_2023} and to measure internal migration from phone records \cite{blumenstock2012inferring,chi2020general,tai2022mobile}, and illustrates the value of combining both streams of analysis in a single setting.

Our second contribution is to provide insight into the relationship between civil conflict and seasonal labor flows in Afghanistan, a country with a long history of violence. We focus on opium poppy cultivation, a key source of rural employment in Afghanistan over the last 30 years \cite{byrd_drugs_2004, mansfield_state_2016, UNODC2021}. The harvest is labor-intensive and generates substantial demand for temporary work. Opium harvesters (called `lancers') lance poppy capsules during the day, and return the next day to collect the resin that the capsules secrete overnight. This process can be repeated up to six times during the harvest, and the entire harvest process for a single farm takes 8-12 days. These seasonal jobs are well-paid relative to other casual labor in Afghanistan, sometimes paying up to five times the wages for unskilled work in rural areas \cite{byrd_drugs_2004}. The United Nations estimates that between 119,000 and 190,000 full-time workers are employed in this industry in Afghanistan annually \cite{UNODC2018, UNODC2021}; in 2020 alone, the opium poppy economy is estimated to have earned the Taliban close to 400 million USD \cite{felbab-brown_pipe_2021}. 

Our empirical analysis begins by characterizing the seasonal labor flows that occur for the opium poppy harvest. The migration data inferred from mobile phone records are very granular, and make it possible to observe the quantity of migration as it varies over time and space. For instance, we find evidence of a concentrated influx of individuals into districts that cultivate poppy -- roughly 3\% more than occur in non-harvest times -- with no comparable influx into districts that do not cultivate poppy. In total, our rough estimates imply that between 54,000 and 85,000 seasonal migrants participate in the poppy harvest annually (see \textit{Materials and Methods}). Roughly two-thirds of the seasonal migrants return to their origin districts within three months of their arrival.

We then explore the role of civil conflict in mediating these seasonal labor flows. We consider two different types of conflict: 
fatal violent events (which occur on a short time scale), and territorial contestation and control by the Taliban (which evolves more slowly over time). We find that seasonal labor flows are \textit{not} significantly impacted by violent events, including extreme violence resulting in a large number of fatalities. This adds nuance to prior evidence that similar violent events do cause displacement \cite{tai2022mobile,wycoff2023forecasting}, and suggests that labor migration may be more resilient to idiosyncratic violence than other types of internal migration. In the Afghan context, this may occur because opium poppy cultivation has been a mainstay of rural agriculture \cite{mansfield_state_2016} that has a stabilizing effect on the rural economy \cite{goodhand_corrupting_2008, koehler_modes_2022}. We further find that, while seasonal labor flows are not impacted by violent events, these labor flows \textit{are} influenced by Taliban activities: harvest-induced migration is significantly higher into poppy-growing districts where the Taliban are present than into poppy-growing districts without Taliban. As we discuss later, this may relate to broader strategic decisions made by the Taliban to encourage opium poppy cultivation \cite{farrell2005have, kreutzmann_afghanistan_2007, azizi2024nature}.  Finally, we observe that government-led counter-narcotics efforts to eradicate poppy suppress harvest-related migration, but they do not eliminate it entirely. This is consistent with prior evidence documenting the limited effectiveness of eradication programs \cite{mansfield_pain_2008, mansfield_state_2016}.

Taken together, our results shed light on the dynamics of conflict and labor mobility in Afghanistan, providing a granular quantitative complement to more qualitative research on the political economy of poppy cultivation in the rural Afghan economy \cite{ farrell2005have, goodhand_corrupting_2008, mansfield_state_2016, koehler_modes_2022, azizi2024nature}. More broadly, our work illustrates how new data sources can reveal social and economic dynamics that would otherwise be challenging to observe, particularly in regions with active conflict.

\section*{Results}
\label{sec:results}

\subsection*{Observing seasonal labor flows with mobile phone and satellite data}
We use a large database of mobile phone Call Detail Records (CDR) to measure migration over the period from April 2013 to October 2020. The CDR provide pseudonymized metadata on where and when subscribers use their phones. In total, we observe 23.3 billion call records from 11 million subscribers, who connect to 18,549 unique cell towers across Afghanistan --- see \textit{Supplementary Figure S1} for a map of these active towers. The data cover six full spring harvest seasons, the main harvest season for opium poppy in Afghanistan. Using methods developed in Tai et al. \cite{tai2022mobile} to infer movement on a daily level, we can calculate the number of individuals moving between districts over this time period (Afghanistan contains 398 districts in 34 provinces).%
\footnote{\cite{tai2022mobile} use the same mobile phone dataset to estimate the immediate effect of violent conflict on internal displacement. We build on that analysis to study seasonal labor, which occurs over longer time horizons and which is influenced by a different set of factors.}

\begin{figure*}[b!]
    \centering
	\includegraphics[width=\textwidth,trim={0 0 0 0cm},clip,page=1]{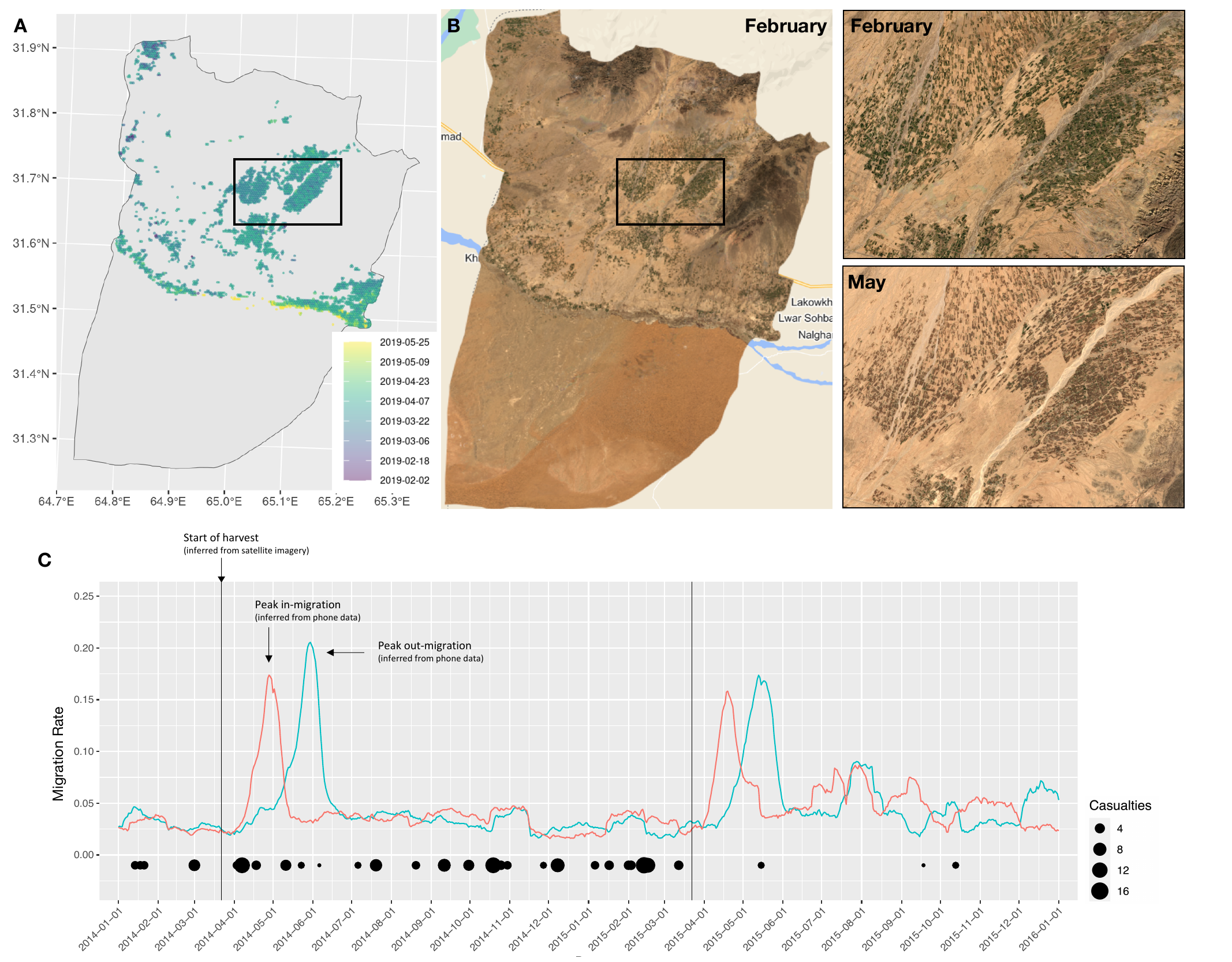}
	\caption[asdf]{\textbf{Using satellite and mobile phone data to study labor migration.} Example of Maywand district in Kandahar Province, a district with high levels of opium poppy cultivation.  \textbf{A}: Map indicates the date when each pixel reaches its peak Normalized Difference Vegetation Index (NDVI) in 2019. \textbf{B}: High-resolution satellite imagery from the Sentinel-2 satellite in February 2019. The enlargements (right-most images) show vegetation in February, before NDVI peaks, and in May, after NDVI peaks. \textbf{C}: The red and blue lines represent the percentage of in- and out-migrants respectively, as measured using mobile phone data (see \textit{Materials and Methods}). Black dots indicate when violence occurred in Maywand, with the size indicating the associated number of casualties.}
	% \includegraphics[width=17.8cm,trim={0 0 0 0cm},clip,page=1]{fig1\_edited.pdf}
	% \caption*{\footnotesize \textit{Notes:} }
	\label{fig:fig1}
\end{figure*}

We then use satellite-based measures of vegetation to infer the approximate timing of harvest in each district in each year. This step is needed because we are not aware of any other systematic dataset tracking these varying harvests across all districts of Afghanistan (see \textit{Materials and Methods} for details).
The methodology is illustrated in Figure~\ref{fig:fig1} for the Maywand district in Afghanistan's southern province of Kandahar. 
Using imagery from NASA's Moderate Resolution Imaging Spectroradiometer (MODIS), we identify the week in which the Normalized Difference Vegetation Index (NDVI), a measure of vegetation density, is at its peak (Figure~\ref{fig:fig1}A-B). This peak typically occurs shortly before harvest \cite{simms2014application}.  Shortly after the date of peak NDVI (the vertical line in Figure~\ref{fig:fig1}C), there is an influx of seasonal workers into the district (the red line), and a corresponding outflux roughly 2-3 weeks later (the blue line).

\subsection*{Estimates of harvest-induced labor migration}
\label{ssec:harvest}

\begin{figure*}[t!]
 % \vspace*{-.5cm}
   \centering
	\includegraphics[width=\textwidth,trim={0 0 0 0cm},clip]{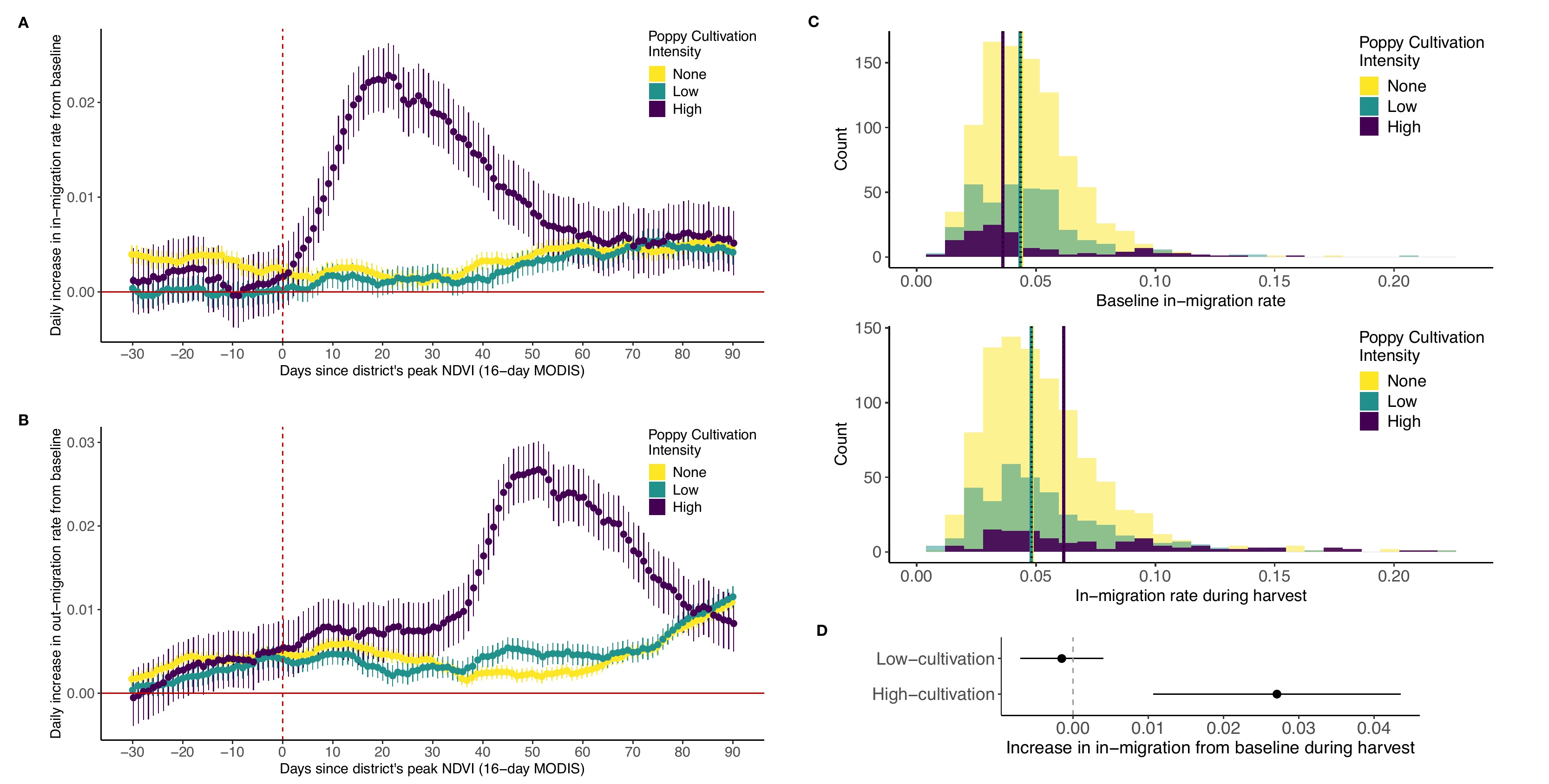}
	\caption{\textbf{Relationship between opium poppy cultivation intensity and seasonal migration.} 
\textbf{A}: Increase in daily in-migration in the 120-day window surrounding the date of peak NDVI ($x=0$). Each dot represents the mean increase in in-migration across district-years with specified cultivation intensities. High cultivation refers to 1000 hectares and above (approximately the top decile of cultivation, according to UNODC district-level estimates); low cultivation is between 1 and 999 hectares. \textbf{B}: Same as \textbf{A}, but now displaying \textit{out}-migration from district-years with specified cultivation intensities.
\textbf{C}: Histograms show the distribution of in-migration rates for district-years with different levels of poppy cultivation intensity at baseline (top) and during the harvest (bottom). Vertical lines represent the median in-migration rate. \textbf{D}: Regression coefficients estimate the increase in harvest in-migration (change in percentage of in-migrants during the harvest compared to the baseline period) for high-cultivation and low-cultivation districts, relative to districts with no poppy cultivation, while controlling for district, province, and time-related characteristics. Bars indicate 95\% confidence intervals. Full regression results are provided in \textit{Supplementary Table S6}.}
	\label{fig:fig2}
\end{figure*}

Using the quasi-random timing of peak agricultural activity to identify seasonal migration, we can compare migration into districts with different types of crop profiles (Figure \ref{fig:fig2}). Most striking are the differences in migration patterns into high poppy-cultivation districts (i.e., the top decile of cultivation, according to district-level estimates from the United Nations Office on Drugs and Crime (UNODC)), relative to districts with low and no poppy cultivation. For instance, Figure \ref{fig:fig2}A indicates that in high-cultivation districts, in-migration is on average over 1\% higher for most of the 45 days after the peak of NDVI than it is during other periods of the year (the peak is 2.3\% on day 21, 95\% C.I.: (1.94, 2.63)\%, $P < 0.001$ for the test that the mean on day 21 is different from zero; this and all subsequent reported statistical tests are two-tailed tests with significance level 0.05). However, districts with no cultivation do not see corresponding changes during that same period. Out-migration from high-cultivation districts begins to increase shortly after in-migration rises, and peaks around day 50 (Figure \ref{fig:fig2}B). There is no corresponding change in out-migration in districts with low or no poppy cultivation.%
\footnote{Note that while our method for inferring peak NDVI is not specific to opium poppy, the clear heterogeneity shown in Figure \ref{fig:fig2}A-B suggests that most seasonal labor flows are associated with opium poppy cultivation; this finding is also consistent with the fact that opium is the most labor-intensive of crops grown in Afghanistan. Full details on our estimation strategy, including a discussion of the fixed effects and time-varying controls, are given in \textit{Materials and Methods}.}

Figure \ref{fig:fig2}C provides further intuition for how patterns of migration differ between baseline (non-harvest) seasons and the harvest season, for districts with different levels of poppy cultivation. In the top (baseline) panel, we observe similar levels of migration across districts with different levels of cultivation -- if anything, there is less migration to high-cultivation districts than other types of districts (i.e., the purple distribution is to the left of the green and yellow distributions). However, during the harvest season (bottom panel), we find that migration rates are substantially higher to high-cultivation districts than to other districts. The figure also illustrates how there is little change in migration rates to low- and no-cultivation districts between baseline and harvest seasons.

Since a variety of factors may contribute to the trends observed in Figure \ref{fig:fig2}A-C, we use a regression specification to estimate the increase in migration into high-cultivation districts during the harvest season, while controlling for district, province, and time-related characteristics (see \textit{Materials and Methods}). The regression coefficients, shown in Figure \ref{fig:fig2}D, imply that on average, high-poppy cultivating districts have a mean daily increase in harvest in-migration of 2.7\% (95\% C.I.: (0.81, 4.61)\%, $P = 0.001$), relative to districts with no poppy cultivation. For this effect to be interpreted as causal, identification relies on the assumption that no other confounding variables affect both poppy cultivation intensity and in-migration at harvest time. We believe this assumption to be plausible since cultivation intensity is determined during the planting season, which occurs several months prior to harvest \cite{tiwari2020wheat}.\footnote{Factors that would violate this assumption would need to influence both the planting season (to affect the amount of cultivation), and the harvest season (to impact in-migration). For instance, if there were a religion that both encouraged poppy cultivation and had important harvest festivals (that induced harvest in-migration), and adherence to that religion varied within a district over time, this could bias our estimates. However, we are not aware of any such confounds.}

What do these coefficients imply about the total influx of seasonal workers to opium producing districts? To answer this question, we multiply the estimated increase of 2.7\% (Figure \ref{fig:fig2}D)
by official district-level population estimates (see \textit{Materials and Methods}). This calculation implies that the 2016 poppy harvest, for example, caused an additional influx ranging from 370 migrants for the smallest high-cultivation district (with a population of roughly 13,500) to 4,000 for the largest (population 145,000). For the six harvest seasons that we analyze, we estimate that, in total, there were roughly 54,000 to 85,000 seasonal migrants annually. However, this is likely a conservative estimate: in sensitivity checks (\textit{Supplementary Table S2}), we find that the estimated migration response is smaller in districts where the dates of peak NDVI are less precise; this suggests that measurement error may attenuate our estimates in districts with less pronounced opium harvests. To our knowledge, no reliable estimates exist on the number of migrant workers hired to work the opium harvest. Anecdotal evidence suggests that seasonal job opportunities for the opium harvest number in the tens of thousands \cite{voa2022}, which is on the same order of magnitude as our findings.  In some years, the UNODC estimates the number of full time (200 days of work per calendar year) jobs for weeding and harvesting. These numbers range from 119,000 to 190,000 \cite{UNODC2018}, but are not directly comparable to our estimates since they focus on full-time employment (and thus also consider non-migrant labor).

The mobile phone data also allow us to observe movement patterns \textit{after} the harvest. For instance, we estimate that roughly 62\% of the migrants who enter districts during the poppy harvest return to their origin districts within 90 days after arriving for the harvest (see \textit{Materials and Methods}).

In analysis described in \textit{Materials and Methods} and presented in the supplementary figures and tables, we conduct a variety of robustness checks to ensure that these results are not unduly influenced by how we construct our measure of migration, how we define peak NDVI (and related measurement challenges), as well as modeling assumptions (e.g., different functional forms of the poppy cultivation variable). Our main results do not change under these alternate specifications. In subsequent analysis, we focus our attention on high-cultivation districts, extending the analysis used to produce Figure \ref{fig:fig2}.

\subsection*{How civil conflict affects labor migration}

Our next set of results explore whether labor migration is affected by (i) salient violent events and (ii) broader patterns of Taliban contestation and control. Since these two types of conflict are correlated (\textit{Supplementary Figure S4}), we test for both effects jointly (see \textit{Materials and Methods}). This analysis relies on two additional publicly-available datasets. The first is a standardized dataset that provides geo-coded information about fatal violent events, collected by the Uppsala Conflict Data Program \cite{Sundberg:2013aa}. These data include information about the location, timing, and estimated number of deaths from discrete violent events.\footnote{We considered alternative sources of publicly-available data on violent events, such as the Armed Conflict Location and Event Data Project (ACLED) \cite{eck2012data} and the Global Terrorism Database (GTD) \cite{lafree2007introducing}. However, ACLED data are only available in Afghanistan from 2017 onward, and the GTD data focus more narrowly on terrorist events.} The second is a dataset that indicates which districts of the country were controlled or contested by the Taliban. The latter dataset is based on military assessments and is published by the Special Investigator General for Afghanistan Reconstruction (SIGAR) \cite{sopko2018qr}.\footnote{The Long War Journal \cite{roggio2020mapping} also provides maps of territorial control, based on SIGAR and media sources, but not in machine-readable format. We are not aware of any publicly-available time-varying source of territorial control data that covers our data period.} This measure is imperfect, since territorial control is not an easy concept to quantify \cite{tao2016hybrid,anders2020territorial}. We rely on SIGAR's indicator of Taliban presence -- which indicates that a district was either controlled, influenced or contested by the Taliban (see \textit{Materials and Methods}). In Afghanistan's conflict, certain regions switched repeatedly between Taliban and government control \cite{jackson2021negotiating}; such regions would be classified as having Taliban presence in our analysis.

Our results indicate that, in districts with high levels of poppy cultivation, there is a nuanced relationship between conflict and migration (Figure \ref{fig:fig3}). In the top panel, we observe that the seasonal increase in in-migration  (relative to non-harvest periods) is highest to districts with Taliban presence, and that the seasonal increase is lowest to districts without Taliban presence or violence. The bottom panel shows the regression coefficients associated with these effects, after controlling for time and province fixed effects, and other control variables (\textit{Materials and Methods}).
In high-cultivation districts, those with violence and Taliban presence experience 3.4\% (95\% C.I.: (1.42, 5.29)\%, $P < 0.001$) larger increases in harvest in-migration, relative to the reference group with no poppy cultivation and no conflict, which is similar to what occurs in districts with Taliban presence but no violence (4.2\%; 95\% C.I.: (0.97, 7.34)\%, $P = 0.011$). Districts without Taliban presence see much smaller increases regardless of whether they experience violence or not: 0.86\% for districts with violence (95\% C.I.: (-0.73, 2.45)\%, $P = 0.289$) and 1.0\% for those without (95\% C.I.: (-0.69, 2.65)\%, $P = 0.251$).
The general pattern of these results is robust to a range of sensitivity checks, including when we remove specific regions from the analysis (since districts with the largest cultivation areas also are more likely to have violence; see \textit{Supplementary Figures S5-6}), when we remove specific periods of time from the analysis (in particular, since the data on territorial control does not exist prior to 2017; see \textit{Supplementary Figure S7}), and when we use alternate definitions of violent events (e.g., to confirm that even very extreme violence does not deter migration significantly; see \textit{Supplementary Figure S8}).

The mobile phone data also make it possible to study how conditions at the \textit{origin} affect decisions about seasonal migration. In baseline (non-harvest) periods, we observe that roughly 90\% of migrants to high-cultivation districts come from districts with low or no poppy cultivation (Figure~\ref{fig:fig4}A, top barplot) --- this is perhaps not surprising since we define high-cultivation districts as roughly the top decile of opium cultivation. However, during harvest seasons, this composition changes, as a greater number of in-migrants come from high-cultivation districts (Figure~\ref{fig:fig4}A, bottom barplot). In Figure~\ref{fig:fig4}B, we estimate that the magnitude of this increase, after controlling for  district, province, and time-related characteristics, is  4.7\% (95\% C.I.: (1.72, 7.74)\%, $P = 0.002$) more for high-cultivation destinations than for destinations with no and low cultivation. We do not see a comparable increase for low-cultivation origin districts (95\% C.I. for the difference between the high and low coefficient estimates in Figure~\ref{fig:fig4}B: (0.012, 3.00)\%; $P = 0.033$). (Note that due to different harvest dates, traveling from a high-cultivation to another high-cultivation district does not preclude individuals from participating in the harvest in their origin districts).

While seasonal migrants are more likely to come from high-cultivation districts, they do not appear to be influenced by the conditions of conflict in their district of origin. In particular, Figure \ref{fig:fig4}C disaggregates high-cultivation sources by whether or not they experienced violent events or had Taliban presence.  While all of the coefficients are positive (indicating that migrants tend to come from high-cultivation districts), none of the coefficients are statistically different from each other, and the relative ordering of effect sizes is sensitive to different modeling assumptions (\textit{Supplementary Figure S9}).

\subsection*{Taliban presence and government eradication}

The preceding results indicate that opium harvest in-migration is highest into districts with Taliban presence, but that such migration is not significantly influenced by the conditions of conflict in origin districts. Several potential mechanisms could underlie this effect: the Taliban might strategically enter districts that are projected to have large poppy harvests (and therefore large influxes of seasonal migrants); Taliban presence might result in stronger financial incentives to migrant laborers \cite{newHum2007}; and the Taliban, who regulate travel in districts they control \cite{skaine2010women}, might protect poppy fields and drug trade routes or encourage fighters and other populations to work the harvest \cite{bureau2007,azizi2024nature,gons2012challenges} (though if this last mechanism prevailed, we might expect Taliban presence at the origin to correlate with migration, which we do not observe in Figure~\ref{fig:fig4}C).

While we cannot directly adjudicate between these mechanisms with our data, we do find empirical support for one factor that likely contributes to the correlation between Taliban presence and higher rates of in-migration. In particular, using data from UNODC on government-led and state-enforced efforts to eradicate poppy, we find that harvest migration is negatively correlated with government-led counternarcotics efforts to eradicate poppy: in Figure~\ref{fig:fig5}A, the coefficient of the interaction term between opium cultivation intensity and the percentage of eradication (hectares eradicated as a percentage of total poppy cultivated) is negative and significant (95\% C.I.: (-0.13, -0.00063); $P = 0.048$). Such eradication efforts are enforced primarily by provincial governors and are monitored by the UNODC \cite{UNODC2016}, and entail physically destroying the crop through aerial or ground-based herbicides. 
Eradication is thus limited by insurgent presence: among high-cultivation districts with Taliban presence, 27.4\% have some eradication, compared to 44.4\% in high-cultivation districts without Taliban presence. The stark difference in eradication between districts with and without Taliban presence is also visible in Figure~\ref{fig:fig5}B --- for example, all districts with eradication percentages above 5\% do not have Taliban presence.

\section*{Discussion}
\label{sec:discussion}

Taken together, our analysis provides a nuanced perspective on seasonal labor flows in Afghanistan. Empirically, the combination of satellite imagery and mobile phone data make it possible to quantify seasonal migration at a level of spatial and temporal precision that would be difficult with traditional data, particularly in a region of active conflict. As the UNODC has noted, ``while area under cultivation and locations of opium cultivation are well understood, one of the most persistent gaps in knowledge of the phenomenon has been the lack of systematic information about the number of households and individuals involved or profiting from the trade in opiates in Afghanistan'' \cite{UNODC2021}. Using these non-traditional data sources, we estimate that the annual poppy harvest increases rates of migration by roughly 3\%, and that the poppy harvest leads to roughly 54,000 to 85,000 seasonal migrants annually.

More interesting, perhaps, are the results related to civil conflict. Broadly, we observe that migrants who work the opium poppy harvest are remarkably resilient to the civil conflict surrounding them.  The idiosyncratic occurrence of violence -- including very salient events with a large number of fatalities -- does not significantly impact seasonal migration, whether the violence occurs at home or in the destination. That we do not find a significant impact of violence on labor flows is surprising, since prior work has found that violence can disrupt economic activity \cite{blattman_civil_2010,de_groot_global_2022}, reduce labor supply and demand \cite{ksoll2021guns}, and cause significant internal displacement \cite{Melander:2007aa,engel2007displacement,schon_motivation_2019, tai2022mobile}. In Afghanistan specifically, there is evidence that violence disrupts economic activity \cite{galdo_conflict_2020,blumenstock2018insecurity} and impacts internal (though not necessarily seasonal) migration \cite{Migration:2019aa,tai2022mobile}.%
\footnote{Indeed, Tai et al. \cite{tai2022mobile} find, using the same mobile phone dataset, that violent events have a significant impact on internal displacement in Afghanistan. However, their focus is different: whereas we find that violent events do not significantly disrupt a specific type of (longer-term, seasonal) migration related to the agricultural harvests, Tai et al. focus on the immediate (daily) displacement effects of violence, across the country and unrelated to agricultural harvests.}

There are several possible explanations for why we do not find evidence that violence significantly reduces migration. One is that opium poppy cultivation can offer a ``vital lifeline'' to rural households in Afghanistan \cite{koehler_modes_2022}, so the decision to participate in seasonal labor may not be easily influenced by changing circumstances at the destination. For many households, income from wage labor far exceeds farm income, creating strong incentives to participate in the annual harvest, despite the threat of violence \cite{byrd_drugs_2004}. Next, there might be some degree of habituation towards violence. In other settings, scholars have documented the behavioral changes that can occur as people acclimate to conflict, such as a reduced response to the sounds of gunfire or public warning systems \cite{VanDijcke2023public,allen2008getting,Choudhury2014narco}. The conflict in Afghanistan has been ongoing for decades, so it is likely that behavioral responses to idiosyncratic violent events are more muted than they would be in a new conflict. 
Finally, it is possible that our measure of district-level violence is too coarse to detect the types of events that are most disruptive to migration. In particular, violence along transportation routes \cite{alfano_spatial_2022} or close to poppy fields may more directly impede travel in the same way that checkpoints, roadblocks and other barriers restrict labor mobility \cite{melnikov2020gangs,ksoll2021guns,cali2018labor,mansour2010effects}. In \textit{Supplementary Materials Section 3}, we present suggestive evidence that violence on road networks that border high-cultivation districts with Taliban presence may indeed reduce in-migration. This is consistent with the idea that violence which disrupts travel might reduce migration, even if regional violence in the destination does not.

Whereas seasonal migration is not significantly impacted by regional violence at the destination, it does appear to be sensitive to broader dynamics of territorial control, with migrants drawn disproportionately to areas where the Taliban are present. This is perhaps not surprising, given that the Taliban encouraged and facilitated opium production for strategic reasons during the period we study.\footnote{In contrast, the Taliban has repeatedly banned poppy cultivation when ruling the country --- including in 2021, shortly after our data ends --- for both religious \cite{farrell2005have} and sociopolitical reasons \cite{felbab-brown_pipe_2021,kreutzmann_afghanistan_2007}.} During this time, opium trafficking helped finance the insurgency, increased the Taliban's local legitimacy, and strengthened their influence \cite{goodhand_corrupting_2008,koehler_modes_2022}. They issued fatwas that allowed opium cultivation and trafficking, arguing that the export of opium to western countries and non-Muslims was a permissible act of Jihad \cite{azizi2024nature}. They also protected poppy fields from government-led eradication efforts \cite{azizi2024nature}, and were perceived to offer more stable governance in many parts of the country \cite{koehler_modes_2022}, which might have made districts with Taliban presence more attractive destinations for seasonal migrants.

The fact that we observe qualitatively different effects on seasonal migration for idiosyncratic violent events than we do for broader patterns of Taliban territorial control is consistent with other recent analysis of the Afghan context in this period \cite{galdo_conflict_2020,blumenstock2018insecurity}. For instance, Galdo et al. \cite{galdo_conflict_2020} find that, between 2012-2016, formal economic activity is negatively affected immediately after violence occurs, but that illicit economic activity either increases (in the quarters following violence) or stays the same (immediately after violence).  However, our results on territorial control are different from that documented in prior work in El Salvador \cite{melnikov2020gangs}, which find that mobility restrictions associated with control by non-state actors tends to reduce labor supply, and recent work in the West Bank that finds no effect of barriers and checkpoints on labor supply \cite{cali2018labor}. Of course, the Afghan context is very different, and the context of the poppy harvest is unique to Afghanistan.

Beyond the insights specific to Afghanistan, we hope that the general methodological framework we have developed can prove useful in other regions and contexts. In particular, the combination of satellite data -- which can reveal the timing of crop harvests -- with mobile phone data -- which can provide granular insight into patterns of  internal migration -- may be useful in understanding the determinants and consequences of seasonal migration.

However, this methodology has certain limitations, several of which are discussed in more detail in \textit{Materials and Methods}. Some of these limitations are inherent to the data -- for instance, since pseudonymized mobile phone metadata do not explicitly indicate the motives behind individuals' migration decisions, we are wary of speculating about why people choose to migrate. Other issues are, to varying degrees, addressed by our empirical analysis and robustness tests. For instance, while we cannot differentiate seasonal workers from others who migrate during the harvest season, our harvest analysis focuses specifically on individuals who in-migrate shortly after peak NDVI in order to increase the likelihood that the migrants are involved in the harvest. We believe a third class of limitations can likely be addressed in future work. For instance, while our analysis of satellite data simply looks for peak vegetation as an indicator of the likely precursor to the opium harvest (which we believe is reasonable in Afghanistan, since opium poppy is by far the most labor intensive crop to harvest), more sophisticated computer vision algorithms could likely directly differentiate opium poppy from other crops.

To conclude, understanding the extent to which conflict affects seasonal labor flows is one step towards understanding the broader effects of conflict, its dynamics, and how to promote peace and post-conflict recovery. We hope that this work illustrates the potential of using non-traditional data sources to study behavior in conflict settings, and encourages additional research in understanding the burden of armed conflict as well as potential strategies to mitigate its impact.

\section*{Materials and Methods}
\subsection*{Data}
\subsubsection*{Satellite imagery}
\label{ssec:satellite}

To measure agricultural activity, we use a standard vegetation index computed from measurements by the National Aeronautics and Space Administration (NASA)'s Moderate Resolution Imaging Spectroradiometer (MODIS) instrument. Data are generated every 16 days at 250 meter spatial resolution worldwide, starting in 2000, and are publicly available.\footnote{https://developers.google.com/earth-engine/datasets/catalog/MODIS\_061\_MOD13Q1} This data product calculates the Normalized Difference Vegetation Index (NDVI), which indicates the density of ``greenness'' in each pixel. NDVI values range from -1 to 1, with clouds and snow having negative values, urban areas having values close to zero, and areas with dense vegetation canopies having positive values (approximately 0.3 to 0.8). Afghanistan's land area is composed of approximately 11 million MODIS pixels, with each pixel having NDVI values for 23 16-day periods each year.

\subsubsection*{Call Detail Records (CDR)}
\label{ssec:phoneData}

To measure patterns of migration, we use a large dataset of pseudonymized mobile phone metadata from one of Afghanistan's largest mobile phone operators, which covers the period from April 2013 to October 2020. These Call Detail Records (CDR) provide metadata on billions of phone calls and text messages.  For each such network transaction, we observe a hash for the subscriber, as well as a date, time, and the cell towers through which the call was routed. Information about the locations of cell towers allows us to approximately locate each subscriber at the time of the transaction -- to within roughly 500 meters in urban areas and roughly 10km in rural areas. The spatial distribution of cell towers and a full description of methods to infer subscriber locations from call detail records can be found in Tai et al. \cite{tai2022mobile}. In the period of our study, there were 18,549 unique cell towers. Grouping towers that are close (less than 100 meters) in physical proximity, we identify 1,795 tower groups. The locations of these cell tower groups are shown in \textit{Supplementary Figure S1}. We are missing data from March and April 2017, two months that coincide with the harvest season, so we limit our analysis to six harvest seasons: 2014, 2015, 2016, 2018, 2019 and 2020.

\subsubsection*{Opium poppy cultivation and eradication}
Data on opium poppy cultivation are collected annually by the United Nations Office of Drugs and Crime (UNODC), aggregated at the district level, and released publicly \cite{UNODC2021}. These data are available from 2003 to 2021 and are collected based on field surveys and manual annotation of high-resolution satellite imagery. We use data from 2014-2020, which overlaps with the period for which phone data are available. In our main analysis, we classify districts by their intensity of cultivation, denoting a district as ``high cultivation'' if its annual cultivation exceeds 1000 hectares, which corresponds to approximately the top decile of districts. A district is classified as ``low cultivation'' if its cultivation is 1-999 hectares, and ``no cultivation'' if 0 hectares. 

Between 2014 and 2016, the UNODC also collected and disseminated data on the number of hectares of poppy that were eradicated \cite{UNODC2016}. In this period, 21.6\% of all districts that had cultivation (and 32.5\% of high-cultivation districts) recorded some eradication. For 2018 and 2019, eradication data are only available at a province level; in 2020 no verified data are available. %In 2014, 2015, 2016, 2018 and 2019, the total number of hectares eradicated were 2692, 3760, 355, 406, and 21. 
For the analysis on eradication (see \textit{Eradication}), we limit our period of study to 2014-2016, where district-level eradication data are available.

\subsubsection*{Violent events} 
Data on violent events are obtained from the Uppsala Conflict Data Program \cite{Sundberg:2013aa}.\footnote{UCDP Georeferenced Event Dataset (GED) Global version 21.1 (\url{https://ucdp.uu.se/downloads/})} This is an open-source collection of data on armed conflict and organized violence, and is widely used in empirical conflict research. The data are derived from media reports; the criteria for inclusion are that the violence must be attributable to an organized actor and result in at least one recorded death. Data are available from 1989 to the present day. 
For each recorded violent event, information on the specific location and temporal duration are also available, with an associated level of precision. Between 2013 to 2020, 18,421 events were recorded in Afghanistan 
in which the district and week are known. We discard 18,458 events that are recorded with less spatial or temporal precision.

\subsubsection*{Territorial control}
Territorial control data are publicly available from the Special Investigator General for Afghanistan Reconstruction (SIGAR)'s 2018 Quarterly Report.\footnote{\url{https://www.sigar.mil/pdf/quarterlyreports/Addendum\_2018-01-30qr.pdf}} Districts were assessed by the NATO-led Resolute Mission on the dimensions of governance, security, infrastructure, economy, and communications, and then classified as being under insurgent control, insurgent influence, contested, under Afghan government influence, or Afghan government control. While assessments were made quarterly starting from late 2015, district level data were made publicly available only for October 2017. We use the levels of contested, insurgent influence or insurgent control, to indicate that insurgents have some level of presence in the district. We refer to this as ``Taliban presence'' in the text. Although this SIGAR assessment was made during and not before our entire data period, reports indicate that these military assessments severely understate insurgent control \cite{roggio2016}. Districts that fall under Taliban influence rarely returned to government control during our study period.\footnote{Visually examining the 11 maps in \cite{roggio2020mapping} that overlap with our data, from 10/1/17 to 7/8/20, we find that there only four cases of districts returning to government control.} We are not aware of other time-varying measures of territorial control covering our study period, which are also publicly available.

\subsection*{Measuring harvest migration}
\subsubsection*{Peak agricultural activity}
We use remote sensing data to infer the timing of agricultural seasons in Afghanistan. This may seem unnecessary, since in principle people on the ground, and organizations monitoring agriculture -- such as the UNODC \cite{UNODC2021}, the Ministry of Agriculture Irrigation and Livestock \cite{tiwari2020wheat} and Alcis (\url{https://www.alcis.org/poppy}) -- are likely to know the timing of many harvests in particular regions. However, we are not aware of any systematic or publicly available data source that provides such information in a consistent and comprehensive manner.

We therefore use MODIS data to calculate the timing of peak agricultural activity in each district in each year, which we refer to as the ``harvest'' season for that district-year. We focus on the Spring harvest season, which is much larger than the Fall harvest, and which is the one tracked by the UNODC \cite{UNODC2021}. Only considering dates in the first half of the year (twelve 16-day periods), and pixels classified as agriculture in a publicly available land cover dataset,\footnote{Copernicus land cover dataset, available yearly at 100 meter resolution for 2015-2019 at \url{https://developers.google.com/earth-engine/datasets/catalog/COPERNICUS\_Landcover\_100m\_Proba-V-C3\_Global?hl=en\#bands}. Category 40 denotes agricultural pixels.}
we determine the maximum NDVI value for each pixel for each season. Then, we identify the 16-day window that is most frequently identified as the maximum, across all pixels exceeding a maximum NDVI of 0.3 (corresponding to likely vegetation) in each district. We refer to the start of this 16-day window as the district's ``peak NDVI'' date. %The underlying assumption is that the peak NDVI occurs roughly just before the harvest. 
Note that this step does not distinguish between opium poppy and other crops -- see the section \textit{Validation of NDVI-based estimate of harvest} for further validation of this measure of activity.

\subsubsection*{Measuring migration into and out of districts}
We extract subscribers' daily ``residence'' from CDR using the same methodology that was developed and validated in \cite{tai2022mobile}. Briefly, we first infer a subscriber's daily district to be the modal district over each 24-hour period. Then, we infer a longer-term ``residence,'' roughly corresponding to a subscriber being in a district for at least a week-long period, using the methodology in \cite{chi2020general}. For each individual, this algorithm returns location segments that cover a specified minimum number of days. Based on these individual location segments, we construct a variety of migration-related measures for each district $d$ at time (day) $t$. 

In-migrants are defined as subscribers in district $d$ on day $t$, that were known to be in a different district 30 days ago, i.e., had location segment in a different district 30 days prior. The proportion of ``in-migrants'', used to measure in-flows on each day $t$, is calculated as the number of subscribers who move to district $d$ on day $t$ divided by the number of subscribers in district $d$ on day $t$ (including the movers). Similarly, the proportion of ``out-migrants'' from $d$ on $t$  is calculated as the number of subscribers who were in $d$ 30 days before $t$ ($t-30$), but not in $d$ on $t$, divided by the number of subscribers in $d$ on $t-30$.
Examples of these measures are in Figure \ref{fig:fig1}C. We discard values of 0 and 1 (2\% of all observations), which are likely due to data sparsity (it is unlikely that all or none of the subscribers present on any day are migrants).%
\footnote{In robustness checks, we confirm that our main results on in-migration are qualitatively unchanged if we use different reference periods to define in-migration, e.g., if  in-migrants are defined as subscribers who were in district $d$ on day $t$, who were known to be in a different district 15 or 45 days ago --- see \textit{Robustness Checks} and \textit{Supplementary Table S1}. 
}

When analyzing the source districts of in-migrants (Figure~\ref{fig:fig4}), we disaggregate in-migrants based on characteristics of the district from which they came (their `home' district), based on the level of poppy cultivation in their home district, Taliban presence in their home district, and whether their home district experienced a violent event in the 30 days prior to $t$. The use of the 30-day window (i.e., comparing locations 30 days ago to today) increases the likelihood that the previous location captured is a subscriber's ``home'' location, rather than a district that they are passing through or have stopped by prior to their eventual destination. 

\subsubsection*{Migration during baseline and harvest seasons}
We measure daily changes in migration-related variables (extracted from the CDR) during the period surrounding peak agricultural activity (extracted from satellite imagery). We define the ``baseline'' period to be between 31 and 120 days before the district's peak NDVI date $t_0$ (see above), and compute the mean value of migration during this period, $M_{dy}^{Base}=\frac{1}{90} \sum_{t = t_0 - 120}^{t_0 - 31} M_{dt}$, where $d$ indexes districts, $y$ indexes years, and $t$ indexes days. 
We then compute the increase in migration (relative to baseline) for each district-day as $E_{dt} = M_{dt} - M_{dy}^{Base}$.
District-years in which fewer than 90 observations are available during the baseline period are dropped from the analysis (5.3\% of all observations).

\label{sssec:excessHarvestMig}
For each district-year, we calculate the average daily increase in migration that occurs during the harvest season. Based on the typical timing of the poppy harvest relative to peak NDVI \cite{simms2014application}, we focus on the period starting between 15 and 35 days after peak NDVI --- this is also the time when in-migration is highest to high-cultivation regions (Figure \ref{fig:fig2}A). Since poppy lancers typically work for 8-12 days on a plot of land \cite{UNODC2021}, we define the increase in harvest migration $E_{dy}$ (for each district-year) to be the 7-day period with the highest mean migration, across all possible 7-day windows during this period. Formally, $E_{dy} = \frac{1}{7} \max_{i \in [15, 29]} \sum_{t = t_0 + i}^{t_0 +i + 6} E_{dt}$. This district-year measure of migration is the main outcome variable in subsequent regressions. In robustness checks, we show that results do not change much if we use different periods of time to search for high levels of migration (i.e., days 1-45 instead of 15-35), or if we use a different number of days to compute the mean (i.e., 14 instead of 7)  --- see \textit{Robustness checks} and \textit{Supplementary Table S1}. A figure showing differences in migration during the harvest and baseline periods, between all pairs of districts in 2015, is provided in \textit{Supplementary Figure S12.}

\subsection*{Estimation framework}
Above, we have discussed how the mobile phone and satellite data make it possible to measure the flows of people into and out of districts in the period of time surrounding the poppy harvest. Figure \ref{fig:fig2}A-C illustrate these measures. To better isolate the migration effect of the poppy harvest, the timing of which varies from year to year and district to district, while controlling for other factors that relate to migration, we use a regression framework.

Our base specification regresses $E_{dy}$, the increase in migration during harvest (relative to the baseline period), on measures of poppy cultivation and a range of control variables, 
\begin{equation}
    \label{eq:mainSpec}
    E_{dy} =  \boldsymbol{\beta_{p}}f(\text{poppy}_{dy}) +  \boldsymbol{\beta} \mathbf{X_{dy}} +   \gamma_p  + \lambda_y +  \epsilon_{dy}
\end{equation}

where $\text{poppy}_{dy}$ is the level of poppy cultivation in district $d$ in year $y$, as estimated by the UNODC. $f(.)$ allows for a flexible functional form for the poppy cultivation area, including both categorical measures (as described in \textit{Data}) and continuous measures of cultivation (\textit{Supplementary Figure S5}). $\gamma_p$ are province-level fixed effects, which control for unobserved province-level characteristics that may affect both the amount of opium poppy cultivated and patterns of migration. $\lambda_y$ are year fixed effects that control for common effects over time.%
\footnote{We do not include district fixed effects because there is insufficient within-district variation in poppy cultivation. Of the 280 districts in our sample, only 24 have high levels of cultivation in some years but not in others.}
We also include an array of district-level characteristics, both time-varying and time-invariant, $\mathbf{X_{dy}}$, including the yearly amount of non-poppy cultivation,\footnote{Derived by subtracting poppy cultivation area from total agricultural area (see \textit{Data})} population,\footnote{https://data.humdata.org/dataset/estimated-population-of-afghanistan-2015-2016} road infrastructure,\footnote{https://datacatalog.worldbank.org/search/dataset/0038673} land use,\footnote{https://hub.arcgis.com/datasets/hqfao::afghanistan-urban-areas-district-level-1/} availability of healthcare,\footnote{https://data.humdata.org/dataset/hotosm\_afg\_health\_facilities} ethnic diversity and majority,\footnote{Afghanistan Central Statistical Office} and whether or not a district is a provincial capital. $\epsilon_{dy}$ is the error term. Descriptive statistics for covariates, the outcome variable and other variables of interest are summarized in \textit{Supplementary Tables S4-5}.

Results in Figure \ref{fig:fig2}D show how the increase in harvest migration (relative to baseline) differs between district-years with high levels of poppy cultivation and district-years with low levels of cultivation, relative to district-years with no  cultivation:

\begin{equation}
    \label{eq:mainSpecCat}
    E_{dy} =  \beta_{h}\text{ high}_{dy} + \beta_{l}\text{ low}_{dy}  +  \boldsymbol{\beta} \mathbf{X_{dy}} +   \gamma_p  + \lambda_y +  \epsilon_{dy}
\end{equation}

The main identifying assumption behind Equations \eqref{eq:mainSpec} and \eqref{eq:mainSpecCat} is that there are no unmeasured confounders that affect both the level of opium poppy cultivation in a district-year and changes in migration (relative to baseline) during the harvest season.  In our setting, this assumption is more plausible because of the long time lag between planting (which occurs in October or November) and harvest (which occurs roughly five to six months later), and the fact that we are able to use the phone and satellite data to isolate the influx of migrants specifically during the harvest season (the timing of which varies from year to year and from district to district). We cluster standard errors at the district level. All estimated coefficients and reported statistical tests in the main text that are the result of regression-based estimates are two-sided tests for whether the coefficient is different from zero. Due to the large sample size, they can be interpreted as z-tests. A significance level of 0.05 is used throughout. Full regression results, including sample sizes, are in \textit{Supplementary Table S6}. 

\subsubsection*{Robustness checks}
We check the robustness of our results to the construction of our outcome variable, measurement and data issues, as well as modeling assumptions. Additional checks related to the NDVI-based measure of harvest dates are detailed in \textit{Validation of NDVI-based estimate of harvest}. The regression coefficients from the base specification that is used to produce the results in the main text (Figure~\ref{fig:fig2}D) are shown in column (1) of \textit{Supplementary Table S1}.

Columns (2)-(5) of \textit{Supplementary Table S1}vary parameters in the construction of the migration outcome variable. Column (2) uses a 15-day reference period to define in-migration and column (3) uses a 45-day period (Section \textit{Measuring migration into and out of districts}), both leading to very similar results. In columns (4) and (5), we again find very similar results if we use different time windows to construct the harvest maximum measure of migration. In particular, column (4) uses a longer time period during the harvest to identify the timing of peak migration (days 1-45 instead of 15-35), and column (5) takes the mean over a 14-day period (instead of 7 days).

To test if extreme values in opium cultivation (\textit{Supplementary Figure S5}) are driving results for the ``high-cultivation'' category, column (6) removes district-years with over 5000 hectares of cultivation (roughly corresponding to the top 3\%). To test the sensitivity of our results to the functional form of $f()$ in Equation \eqref{eq:mainSpec}, column (7) shows how migration varies with a continuous (log) measure of opium poppy cultivation.

\subsubsection*{Validation of NDVI-based estimate of harvest}
There are several potential issues with the NDVI-based measure that we employ. First, satellite-derived data can contain inaccuracies due to atmospheric conditions such as illumination and cloud cover, sensor calibration, and viewing angles. Second, NDVI can be influenced by factors such as soil moisture, which may not be related to the stage of crop growth. And finally, the measure we develop detects all agricultural activity and is not specific to opium poppy. 

Our analysis relies on MODIS's 16-day composite product, which identifies a representative NDVI value for each 250m pixel for each 16-day window. That representative value is selected by filtering the data based on the quality, presence of clouds, and the viewing angle of the sensor
\cite{huete2002overview}. This process is intended to reduce noise and produce a more reliable estimate that mitigates concerns related to the raw NDVI measurements.  However, it also implies some loss of information about the shape of the NDVI profile, as well as the exact timing of peak NDVI.

To gauge the extent to which these concerns might influence our results, we compare the peak NDVI data calculated using our method to field-based estimates of the harvest in a handful of districts. \textit{Supplementary Figure S2} displays the estimated peak NDVI date for 10 districts that are most frequently classified as high-growing poppy districts, for the period from 2014 to 2020. Our estimates appear broadly consistent with available information. According to the UNODC \cite{UNODC2016}, the months just before the harvest are March and April in South, East and Western regions, and May, June and July in North and North-eastern regions, with dates beginning later as we move northwards. In \textit{Supplementary Figure S2}, the first set of districts (Zhari, Maywand, Washer, Nahri Sarraj, Lashkar Gah and Khash Rod) are in the South and typically have early harvests. Over the years, we see that these consistently have the earliest inferred peak agricultural activity, in March and April. Bala Murghab district is in the Western region and also has an early harvest. Pachier Agam and Khogayani are in the East, which typically has early harvests, but these districts are at higher elevation -- we see that their inferred peak agricultural activity occurs either at a similar or slightly later time, compared to the first two sets of districts. Finally, Argo district is in the North-East, and consistently has the latest inferred dates of May in every year. All these observations are consistent with expectations. 

In principle, the migration outcome variable that we construct should be to be robust to minor inaccuracies in the inferred harvest date. This is because our ``baseline'' measure of migration takes an average over a 90-day period, from 31 to 120 days before peak NDVI. The ``harvest'' measure likewise considers 7-day periods with the highest in-migration that fall between 15 and 35 days after peak NDVI. To the extent that there is classical measurement error in the harvest date, we expect that this would lead to attenuation bias, i.e., to bias our estimates of in-migration to high-growing districts towards zero. In turn, this would imply that our estimates of the effect of poppy cultivation on migration are conservative. 

We nonetheless perform several tests to determine whether inaccuracies in inferring harvest dates can substantively affect downstream results based on those dates.  In the first exercise, we perturb the peak NDVI dates slightly by adding or subtracting 14 days from the estimated peak NDVI date of every district-year. In \textit{Supplementary Table S2}, we show the results from estimating our mean specification (Equation \eqref{eq:mainSpecCat}) with the peak NDVI date perturbed to be 14 days earlier than estimated (column 1) and 14 days later than estimated (column 2). In both cases, the estimated effect for the high-cultivation group remains positive and significant, and as expected, the estimated magnitudes are slightly smaller than without perturbation (original estimate of 0.0271). 

As a second check on the sensitivity of our results to noise in the satellite-based measurements of the harvest dates, columns (3)-(4) of \textit{Supplementary Table S2} disaggregate our results based on the precision with which the peak NDVI date is measured.  Column (3) includes district-years in which the majority (more than 50\%) of pixels in the district have the same peak NDVI date NDVI; column (4) includes the remaining districts. When peak agricultural activity (opium or otherwise) is measured more precisely, the estimated effect is slightly larger; when  the harvest dates are estimated less precisely, the estimated effect is slightly smaller, but still significant at $p<0.01$.

In the final exercise, we conduct a placebo test to check whether the positive and significant effects that we estimate also occur at times of the year not associated with the harvest. Specifically, for each of 250 iterations, we set the placebo peak NDVI date for each district-year to be a randomly selected 16-day period in the first half of the year (to create a placebo Spring harvest). We then recompute the migration outcome variables with respect to these dates, and re-run the main regression (Equation \eqref{eq:mainSpecCat}). The results, shown in \textit{Supplementary Figure S3}, indicate that all of the 250 placebos produce coefficients substantially smaller than what we estimate using the actual peak NDVI date.

\subsubsection*{Estimate of total number of seasonal migrants}
To estimate the total number of migrants per season, we combine our regression estimates (Figure \ref{fig:fig2}D) with external population data. Specifically, we multiply the estimated increase in in-migration to high-cultivation districts (2.71\%) --- this tells us the proportional influx of migrants to high-cultivation districts, relative to that destination district's baseline population --- by official estimates of each high-cultivation district's population, as recorded in 2015 and 2016 by Afghanistan's Central Statistical Office (available at \url{https://data.humdata.org/dataset/estimated-population-of-afghanistan-2015-2016}). This is a back-of-the-envelope calculation that requires several assumptions. First, we note that the regression estimates come from a model with relatively low explanatory power ($R^2$ of 0.13, \textit{Supplementary Table S1}). This suggests that the data are noisy and confidence intervals will be wide (but does not imply that the estimate of 2.71\% is biased). Next, we assume that the mobile subscribers we observe respond similarly to the poppy harvest as people who do not appear in our data. We further assume that estimates based on districts in our sample (see \textit{Supplementary Figure S1} for geographical coverage of CDR; in 2018, 34 out of 44 high-cultivation districts had cell towers) generalize to districts not in our sample. If, for instance, mobile subscribers were more likely than non-subscribers to migrate for the poppy harvest, then our back-of-the-envelope calculation would over-estimate the total number of migrants per season. Despite these limitations, we include these estimates as they facilitate a quantitative comparison of our results to existing anecdotal evidence on seasonal workers in Afghanistan. 

\subsubsection*{Return of seasonal migrants}

To study the locations of migrants \textit{after} the migration period, we use a similar regression framework as was used to construct \textit{Supplementary Table S1-2}. In \textit{Supplementary Table S3}, the first column uses, for the dependent variable, the proportion of in-migrants who remain in the destination district 30 days later; the coefficient indicates that, relative to baseline periods, roughly 7\% fewer migrants during the harvest season remain in high-cultivation districts than in low and no-cultivation districts. In column (2), we observe that harvest migrants to high-cultivation districts are more likely to return to their district of origin (i.e., where they were immediately prior to migrating) 30 days post-harvest than migrants to low-cultivation districts. Column (3) shows a similar difference persist when we consider returning any day during the 90-day period post-harvest.

What do these coefficients imply about the return patterns of seasonal migrants? During baseline periods in high-cultivation district-years, a median of 46\% of in-migrants return to their previously observed district in the 90-day window after in-migrating.  Applying the regression estimate in column (3) to this median, we estimate that 53\% of the harvest migrants to high-cultivation districts return to their previously observed district after the harvest. However, in-migrants during the harvest season consist of both ``regular'' in-migrants, and seasonal migrants. Assuming movement patterns of regular in-migrants stay the same as at baseline, then differences in return patterns between baseline and harvest can be attributed to seasonal migrants. Thus, as a percentage of the number of seasonal migrants, roughly 62\% return to their previously observed district.

\subsubsection*{The role of conflict}
To investigate the role that conflict plays in seasonal migration, we use measures of idiosyncratic violence and  territorial control. We measure idiosyncratic violence with indicator variables for whether one or more violent events occurred in the month before peak NDVI at the destination district. For territorial control, we use an indicator for whether or not Taliban were present in the destination district. Unlike the violence variables, measures of territorial control are time-invariant (see \textit{Data}). We then allow for the effects measured in Equation \eqref{eq:mainSpec} to vary by the nature of conflict at the destination by interacting measures of poppy cultivation with measures of conflict, i.e.,
\begin{equation}
    E_{dy} =  \boldsymbol{\beta}_{pc} (\text{poppy}_{dy} \times C_{dy})
     +  \boldsymbol{\beta} \mathbf{X_{dy}} + \gamma_p  + \lambda_y +  \epsilon_{dy}
\end{equation}
Since violence and territorial control are correlated (\textit{Supplementary Figure S4}), we include interactions between poppy cultivation, violence and territorial control to estimate their effects jointly. We estimate the following equation, where $H_{dy}$ and $L_{dy}$ are indicator variables for high and low poppy cultivation intensity in district $d$ in year $y$, $V_{dy}$ indicates violence in the destination in the month before peak NDVI, and $T_{d}$ indicates Taliban presence.
\begin{equation}
\label{eq:conflictDest}
     \begin{aligned}
E_{dy} & = \beta_{hvt} H_{dy}*V_{dy}*T_{d} + \beta_{lvt} L_{dy}*V_{dy}*T_{d} \\
 &+ \beta_{hv} H_{dy}*V_{dy}  + \beta_{lv} L_{dy}*V_{dy} \\
 &+ \beta_{ht} H_{dy}*T_{d} + \beta_{lt} L_{dy}*T_{d} \\
 & + \beta_{vt} V_{dy}*T_{d}\\
&+ \beta_{h} H_{dy} + \beta_{l} L_{dy} +    \beta_{v} V_{dy}  +    \beta_{t} T_{d}\\
    & +  \boldsymbol{\beta} \mathbf{X_{dy}} + \gamma_p  + \lambda_y +  \epsilon_{dy}
\end{aligned}
\end{equation}

\begin{figure}[t!]
 % \vspace*{-.5cm}
   \centering
	\includegraphics[width=\linewidth,trim={0 0 0 0cm},clip]{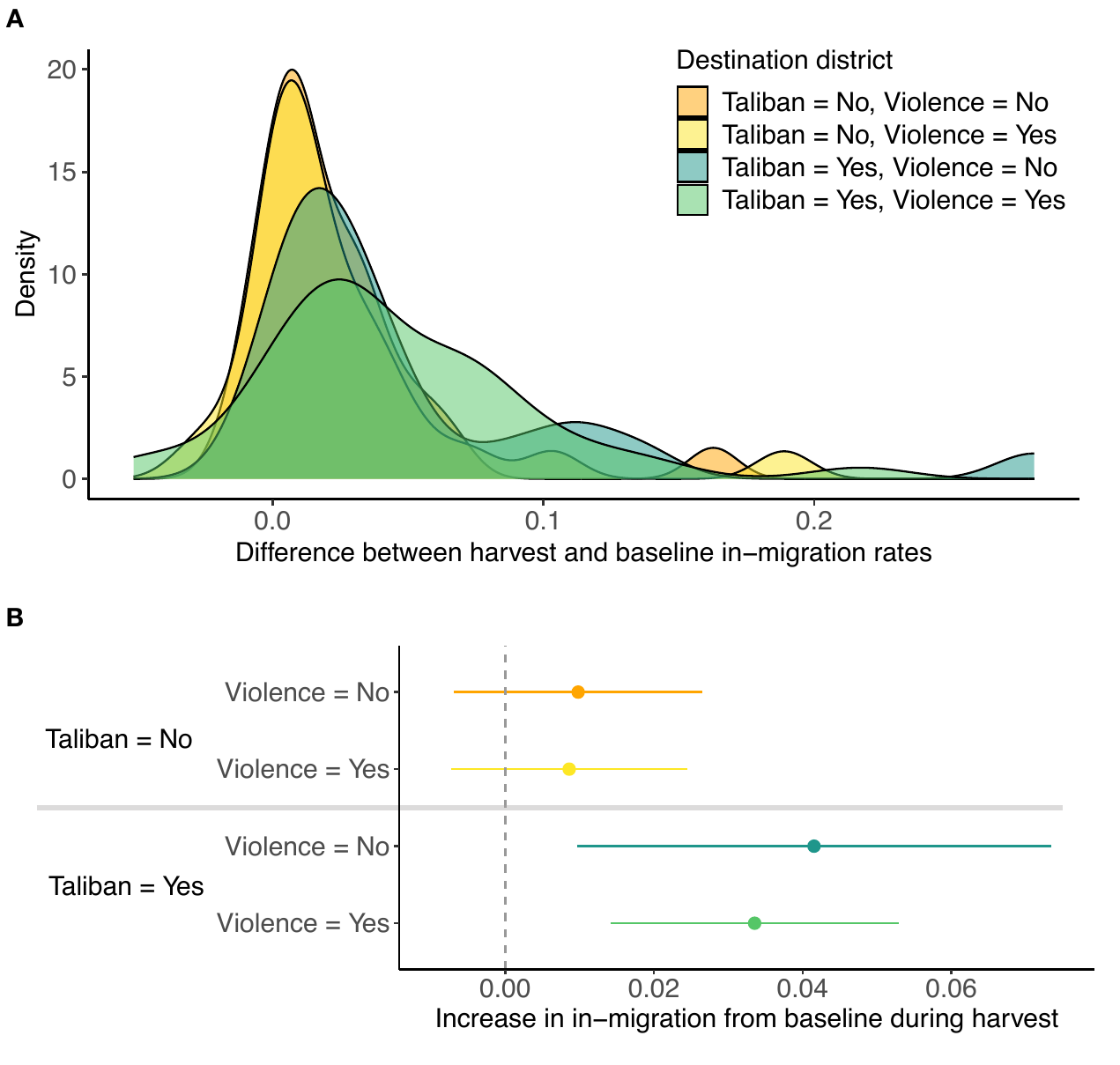}
	\caption{\textbf{Civil conflict and labor migration in high-cultivation districts.} \textbf{A}: Density plots of the increase in harvest in-migration for district-years with high cultivation intensity, for different levels of violence and Taliban presence. \textbf{B}: Regression coefficients estimate the increase in harvest in-migration observed in panel A, relative to districts with no poppy cultivation, no violence and no Taliban presence, controlling for district-level characteristics, unobserved province-level characteristics, and common effects over time. Bars indicate 95\% confidence intervals. Full regression results are provided in \textit{Supplementary Table S6}.}
	\label{fig:fig3}
\end{figure}

\begin{figure}[hp!] 
   \centering
	\includegraphics[width=.9\linewidth,trim={0 0 0 0cm},clip]{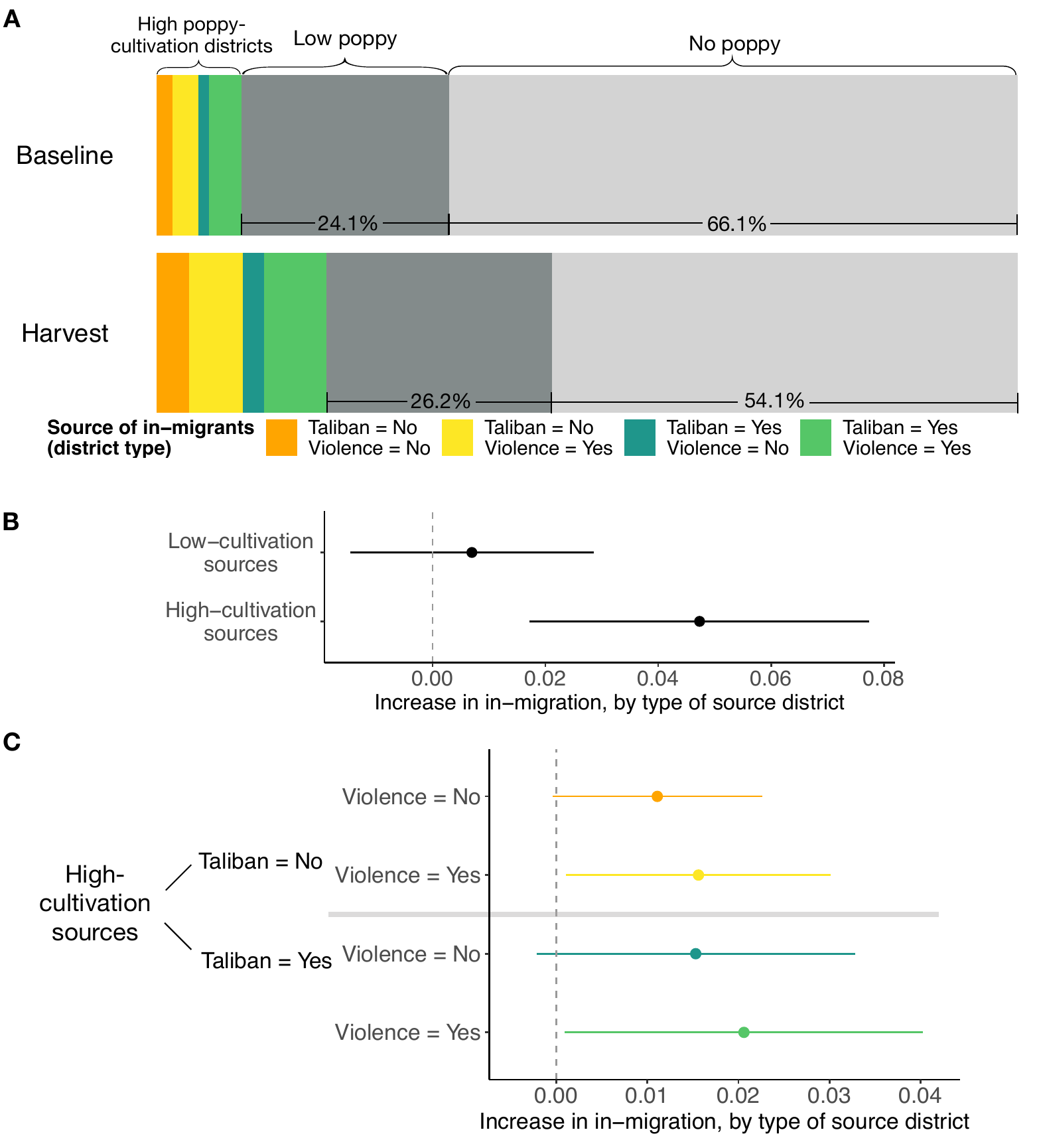}
	\caption{\textbf{Where do seasonal migrants come from?} \textbf{A}: Top barplot indicates the composition (median proportion) of in-migrants to high-cultivation districts during the baseline (non-harvest) period. Bottom barplot indicates the composition during harvest. \textbf{B}: Increase in migrants (during harvest relative to non-harvest, in high-cultivation destinations relative to no-cultivation destinations), by the cultivation intensity of the source district. Regression coefficients (dots) and 95\% confidence intervals (bars) control for district-level characteristics, unobserved province-level characteristics, and yearly fixed effects.  \textbf{C}: Same as \textbf{B}, but now disaggregating by the nature of conflict in the source district. Full regression results are provided in \textit{Supplementary Table S6}.} 
	\label{fig:fig4}
\end{figure}

\begin{figure}[t!]
 % \vspace*{-.5cm}
   \centering
	\includegraphics[width=\linewidth,trim={0 0 0 0cm},clip]{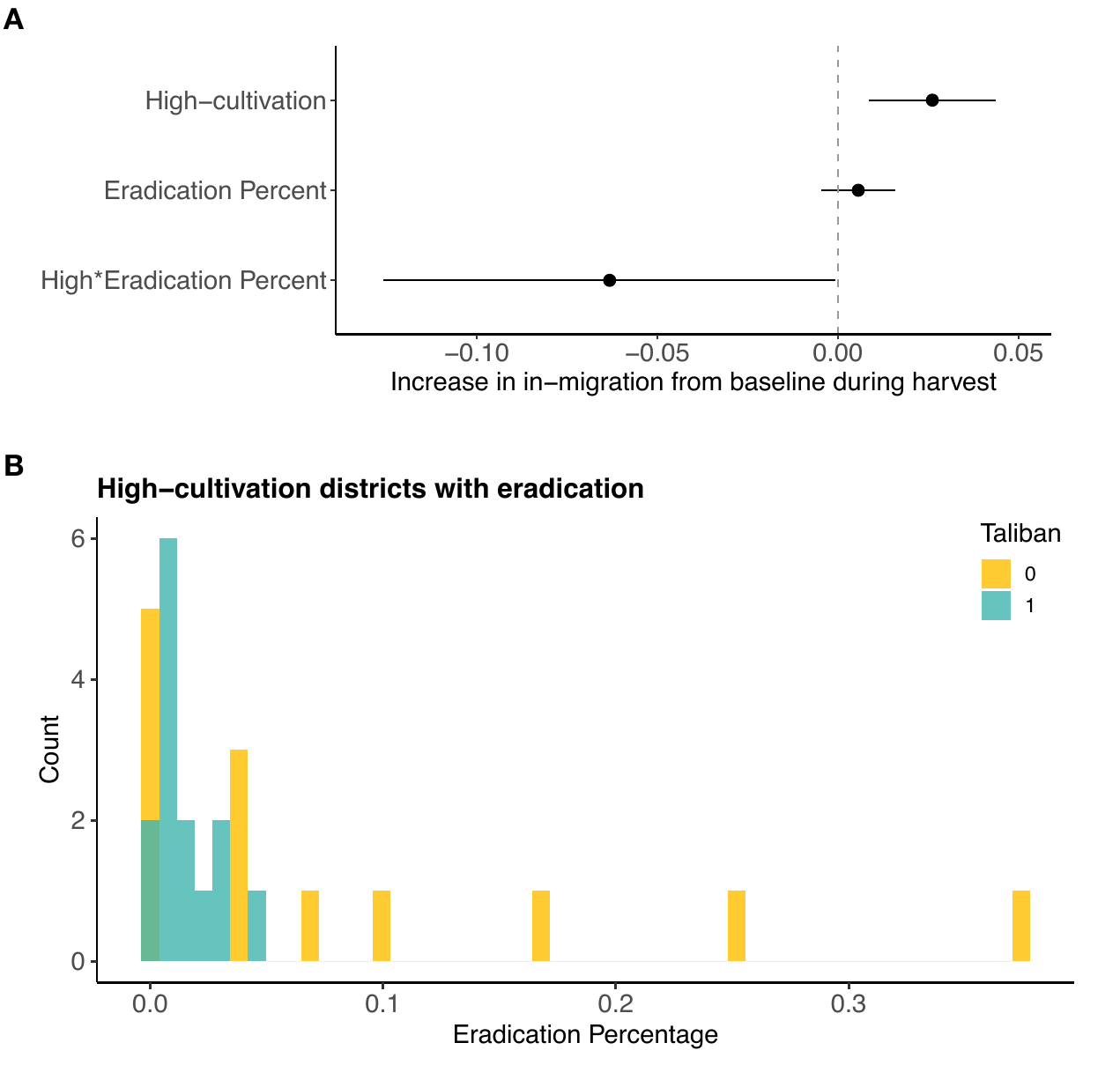}
	\caption{\textbf{Poppy eradication efforts correlate with seasonal migration and Taliban presence.} \textbf{A}: Effect of eradication percentage on the increase in harvest in-migration to high-cultivation districts. Bars are 95\% confidence intervals. Full regression results are provided in \textit{Supplementary Table S6}. \textbf{B}: Histogram of the percentage of total poppy cultivated area eradicated, among high-cultivation districts with any eradication. Colors represent Taliban presence. Only data for 2014-2016 are shown, as district-level eradication data are not available for later years.}
	\label{fig:fig5}
\end{figure}

The coefficients reported in Figure \ref{fig:fig3} represent the estimated difference in the outcome (increase in harvest in-migration) for the groups of interest compared to the reference group (with no poppy cultivation, no violence, and no Taliban presence). The estimate is obtained by summing the relevant regression coefficients, e.g., for high-cultivation districts with violence and Taliban presence, the Figure reports $\hat{\beta}_{hvt} + \hat{\beta}_{hv} + \hat{\beta}_{ht} + \hat{\beta}_{vt} + \hat{\beta}_{h} + \hat{\beta}_{v} + \hat{\beta}_{t}$. (To see this, consider the fitted outcomes for high-cultivation districts with violence and Taliban presence: we have $\hat{E}_{dy} = \hat{\beta}_{hvt} + \hat{\beta}_{hv} + \hat{\beta}_{ht} + \hat{\beta}_{vt} + \hat{\beta}_{h} + \hat{\beta}_{v} + \hat{\beta}_{t} +  \boldsymbol{\hat{\beta}} \mathbf{X_{dy}} + \hat{\gamma}_p  + \hat{\lambda}_y$. For the reference group, we have $\hat{E}_{dy} =  \boldsymbol{\hat{\beta}} \mathbf{X_{dy}} + \hat{\gamma}_p  + \hat{\lambda}_y$. The difference in outcome between the two groups is the sum of coefficients).

In robustness tests, we (1) remove district-years with the highest levels of cultivation, (2) remove individual years of data, (3) use pre-2017 and (4) post-2017 data (\textit{Supplementary Figures S6-7}). We also consider only higher-intensity violence -- more than two events (top decile in terms of number of events), and/or ten or more casualties (top decile in terms of number of casualties) (\textit{Supplementary Figure S8}). 

Our analysis of conflict in source districts is organized analogously to the analysis of conflict at the destination. In Figure~\ref{fig:fig4}B-C and \textit{Supplementary Figure S9}, we show how the composition of in-migrants varies by the level of poppy cultivation at the source and by the nature of conflict in the source (for high-growing destinations) --- all relative to the composition of in-migrants during baseline periods. Estimates are based on Equation \eqref{eq:mainSpecCat}, using outcome variables involving the composition of in-migrants (see \textit{Measuring harvest migration}). For example, the bottom-most plotted coefficient in Figure~\ref{fig:fig4}C uses as the outcome variable the harvest change in proportion of in-migrants that come from districts with high levels of poppy cultivation, experiencing violence, and with Taliban presence.

\subsubsection*{Eradication}
To study the role of eradication in dynamics of migration, we calculate for every district-year the percentage of opium crop eradicated as a fraction of the sum of the number of hectares cultivated and eradicated, based on UNODC estimates. We then interact the categorical poppy cultivation variable in Equation \eqref{eq:mainSpecCat} with a variable that indicates the percentage of eradication at the destination.

\subsubsection*{Limitations}

The methodology that we develop and implement in this paper has several limitations. First, our method of inferring harvest dates from NDVI does not explicitly differentiate between opium poppy and other agricultural crops. In our setting, the opium poppy harvest is uniquely labor-intensive, and relatively short in duration. Thus, while our measurement strategy incorporates information specific to the opium poppy crop (i.e., the timing and length of harvest), it may also be influenced by other similar crops. In our regression analysis, we therefore control for the amount of non-opium cultivation. We find that the relationship between migration and opium poppy cultivation is not sensitive to the inclusion of these controls, and that there is no consistent relationship between non-opium agriculture and migration. A full discussion of the limitations of measuring peak agricultural activity using NDVI, and associated robustness checks, is in the section \textit{Validation of NDVI-based estimate of harvest}.

A related concern arises from the fact that the movement we observe may not consist exclusively of seasonal workers. This limitation is inherent to mobile phone data, which does not directly reveal the motives or occupation of pseudonymized subscribers. For this reason, we are wary to speculate on the factors that draw workers to areas with Taliban presence (e.g., whether it is due to economic incentives or other factors). However, by using the precise expected timing of harvest -- which differs from year to year and district to district -- we focus our analysis on individuals who are very likely to be seasonal workers. For instance, the increases in in-migration that occur in high-cultivation districts immediately after peak agricultural activity, and which last only a few weeks in duration, are unlikely to be Taliban fighters, who typically arrive later (after the harvest) and stay for much longer periods.

More generally, the use of mobile phone data has its own limitations. Chief among these are concerns of personal privacy, particularly in conflict-affected settings and where vulnerable individuals may be involved \cite{salah2019guide,UN2019,vanhoof2018assessing,grassini2021mobile,tiru2014overview,taylor2016no,Blumenstock2018}. We are careful to only analyze pseudonymized data in aggregate form (i.e., at the district level), to ensure that no individual can be re-identified. A separate concern is that the mobile phone data are not necessarily representative of the broader population, and so the inferences we draw should be interpreted as valid for the mobile phone-owning population (Afghanistan's mobile phone penetration rate is currently estimated at 66\% \cite{GSMA2023}) --- to the extent that we extrapolate from one population to the other, we have tried to be explicit about this limitation. In related work, \cite{tai2022mobile} compare broader patterns of internal migration in Afghanistan, inferred from CDR and aggregated at the province level, to estimates of internal displacement from the International Organization for Migration, and find the two measures of migration are correlated with $\rho=0.49$ ($p$=0.004).
A related concern arises from uneven patterns of mobile phone coverage: since we only include districts in our sample if there is at least one phone tower in the district, and since migrants only appear in districts if their phone connects to a tower in the district, this could introduce systematic measurement error into our estimates of migration. For example, if the districts without cell towers are systematically less likely to receive migrants, then we would overestimate the total number of seasonal migrants. Such measurement error could also propagate to our analysis of the impact of violence on migration: if the impact of violence on migration were greater (or lesser) in districts with more cell towers, then our analysis, which relies on cell towers to detect migrants, would over- (or under-) estimate the effect of violence.
Several of these concerns, as well as related issues -- such as the fact that each phone number we observe may not correspond to exactly one individual, and that there are measurement errors in inferring locations from call detail records -- are common to related work using phone data, so we defer to prior work for more in-depth discussion of these issues  \cite{wesolowski2013impact,sekara2019mobile,blumenstock_mobile_2010,tai2022mobile,leo2016socioeconomic}. While it is not possible to eliminate all concerns, we address as many as we can through the data processing, modeling, and sensitivity analysis described in \textit{Materials and Methods}. 

The conflict data that we use also have limitations. The UCDP data set has been used in a large number of research studies, and its potential biases are well-known. In particular, because the data are mainly derived from media sources, they may be biased towards locations with larger media presence, such as urban and more populous areas \cite{Sundberg:2013aa,oberg2011gathering,galtung1965structure}. Likewise, our data on territorial control are from a single point in time; while we would ideally use a time-varying measure of territorial control (which would strengthen a causal argument for the effect of conflict on seasonal labor flows), we are not aware of such data that are currently publicly available.

A final point is that, like other related empirical work on the consequences of conflict, our results are contingent at least in part on aspects specific to the conflict in Afghanistan. The Afghan conflict is endemic and of medium intensity, and results will more easily translate to conflicts with similar characteristics, such as in Colombia, rather than high-intensity conflicts with widespread destruction and displacement. Similarly, our results on seasonal labor are in the context of an illicit and highly lucrative agricultural crop; the broader conclusions that we draw throughout the text reflect this observation.

\section*{Data availability}
With the exception of the mobile phone metadata, all data used in this paper are publicly available and sources have been listed in the text. The mobile phone dataset contains detailed information on over 20 billion mobile phone transactions in Afghanistan. These data contain proprietary and confidential information belonging to a private telecommunications operator, and cannot be publicly released. Upon reasonable request, we can provide information to accredited academic researchers about how to request the proprietary data from the telecommunications operator. With the telecommunication operator's permission, we can also provide district-level aggregate measures of migration for replication purposes to accredited academic researchers.

\section*{Code availability}
All code to reproduce the findings of this study is available at \url{https://github.com/Global-Policy-Lab/afghanistan-civil-conflict-and-labor-flows}.

% Bibliography
\bibliographystyle{naturemag}
\bibliography{displacement,migration,poppy,targeting,afg_qual}

\section*{Acknowledgements}

We thank Jane Esberg, Juan F. Vargas, Christopher W. Blair, and Austin L. Wright for helpful comments and suggestions. Tej Sathe provided excellent research assistance on this project. Blumenstock was supported by the National Science Foundation under CAREER Grant IIS-1942702.

\section*{Author contributions statement}
X.H.T., S.R.N., S.M. and J.E.B. designed and performed research. X.H.T., S.R.N. and S.M. analyzed the data. X.H.T. and J.E.B. wrote the paper. X.H.T., S.R.N. and J.E.B. edited and revised the manuscript.

\section*{Competing interests statement}

The authors declare no competing interest.

\newpage

\section*{Figures}

\clearpage

\renewcommand\thefigure{S\arabic{figure}}
\renewcommand\thetable{S\arabic{table}}
\setcounter{section}{0}
\setcounter{figure}{0}
\setcounter{table}{0}

\section*{Supplementary Materials}
\section{Supplementary Figures}

\begin{figure}[hbtp!]
    \centering
	\includegraphics[width=\linewidth,trim={0 0 0 0cm},clip,page=1]{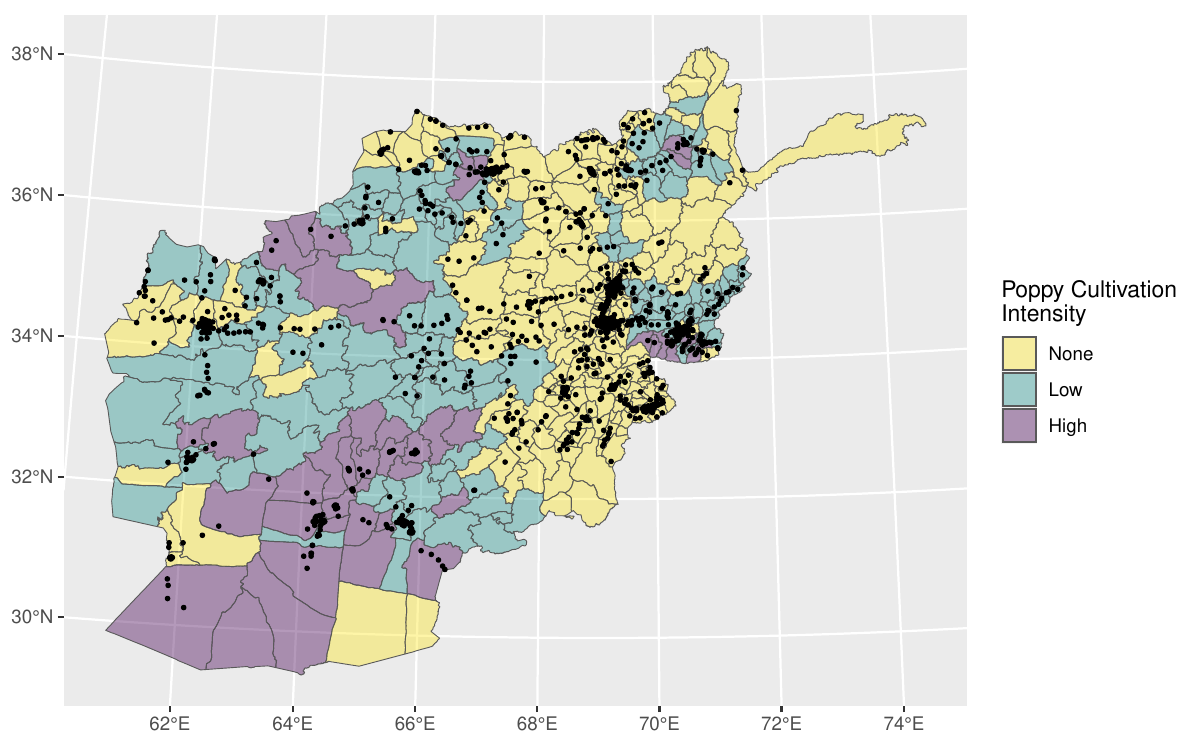}
	\caption{\textbf{Geographical distribution of cell towers and cultivation intensity.} There are 18,549 unique cell towers over the data period (April 2013 to October 2020). Grouping towers that are close (less than 100 meters) in physical proximity, we use information on 1,795 tower groups, represented by black dots. Gray lines are district boundaries, colored by poppy cultivation intensity estimated by the United Nations Office of Drugs and Crime (UNODC) in 2018. We denote more than 1,000 hectares (top decile) as high cultivation, 1-999 hectares as low cultivation, and 0 hectares as no cultivation.}
\end{figure}

\begin{figure}[hbtp!]
    \centering
	\includegraphics[width=.9\linewidth,trim={0 0 0 0cm},clip,page=1]{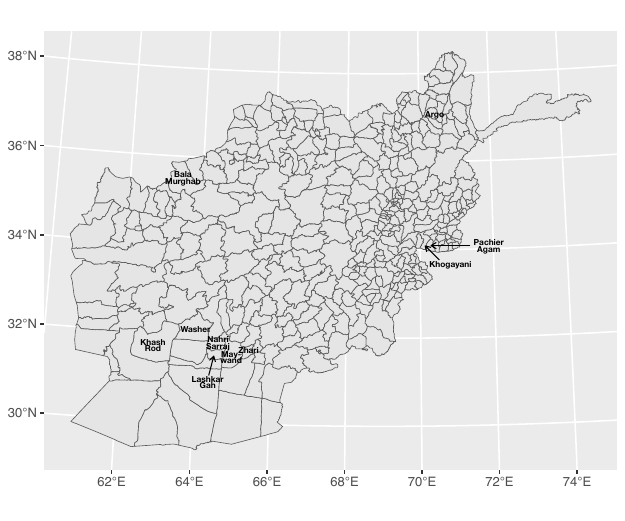}
    \includegraphics[width=\linewidth,trim={20 0 0 0cm},clip,page=1]{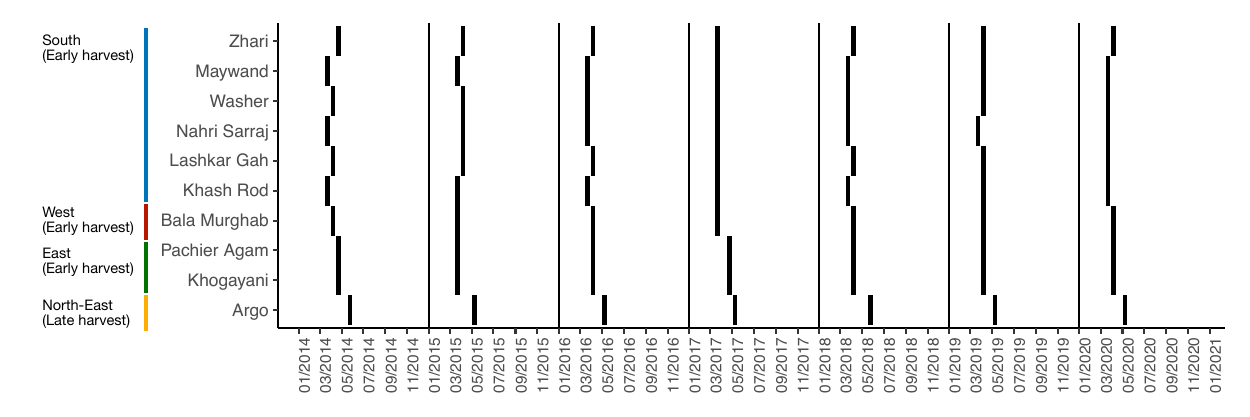}
	\caption{\textbf{Inferred agricultural timelines for districts with high poppy cultivation.} Top: Geographical locations of the 10 districts that are most frequently classified as high-growing poppy districts, for the period from 2014 to 2020. Bottom: Shaded bars display the estimated peak NDVI date for each year from 2014 to 2020, using the methodology described in \textit{Methods, Peak agricultural activity}. Our estimates appear broadly consistent with available qualitative information about the harvest (see \textit{Methods, Validation of NDVI-based estimate of harvest}).
}
\end{figure}

\begin{figure}[hbtp!]
    \centering
	\includegraphics[width=.9\linewidth,trim={0 0 0 0cm},clip,page=1]{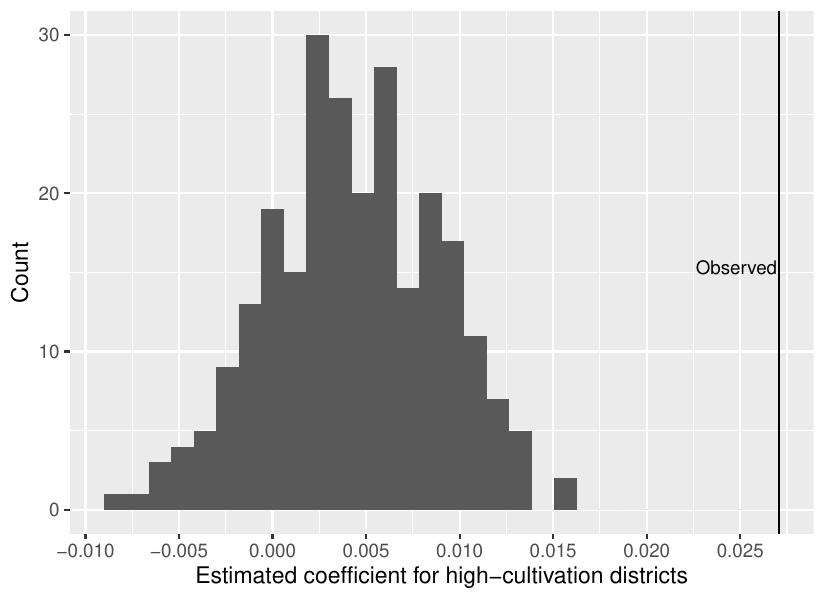}
	\caption{\textbf{Estimated coefficients for placebo test on peak NDVI dates.} Histogram displays estimated coefficients for 250 iterations of the following test: For each iteration, we randomize the peak NDVI dates for each district-year, to be a randomly selected 16-day period (in the first half of the year, to capture the Spring harvest season as in our original methodology). We then recompute the migration outcome variables with respect to these dates, and re-run the main regression (Equation 2). The vertical line (``Observed'') is the estimated coefficient for the original data and presented in the main results. None of the 250 permutations result in a larger coefficient than what we observe in the main results.}
\end{figure}

\begin{figure}[hbtp!]
    \centering
	\includegraphics[width=\linewidth,trim={0 0 0 0cm},clip,page=1]{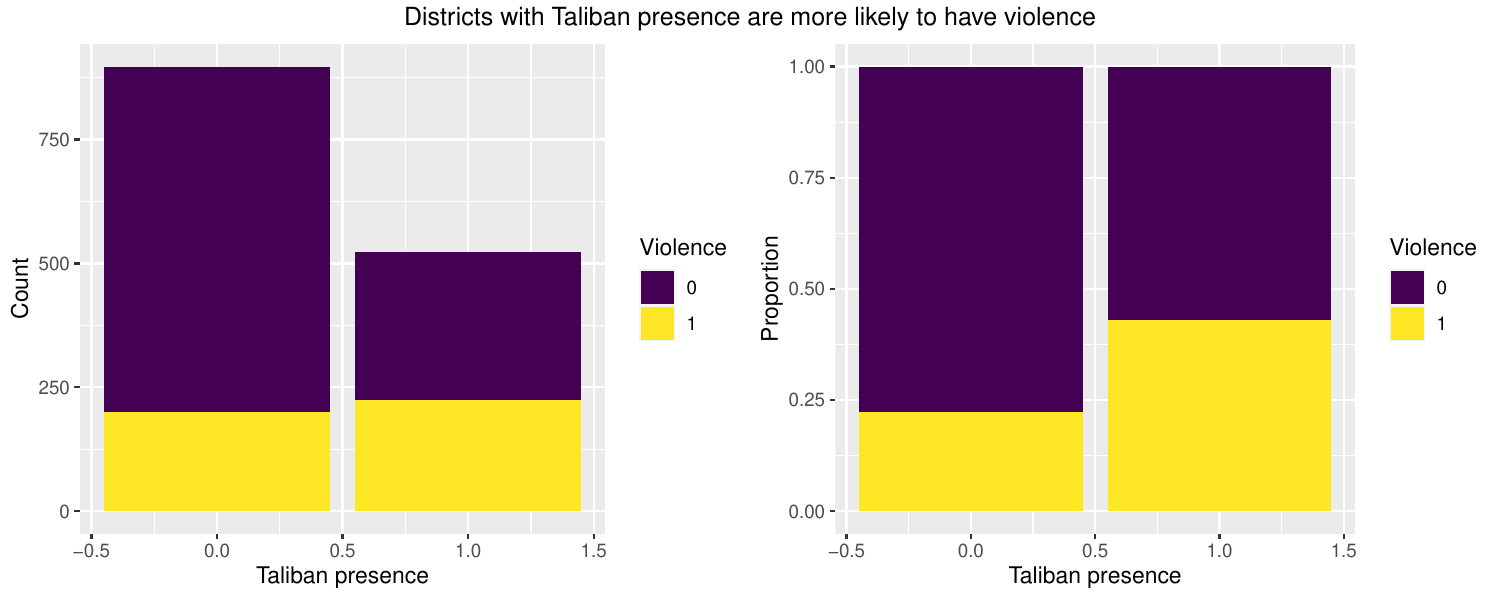}
	\includegraphics[width=\linewidth,trim={0 0 0 0cm},clip,page=2]{9-26-23twoWayEDA.pdf}
	% \caption*{\footnotesize \textit{Notes:}}
	\label{fig:violenceControl}
	\caption{\textbf{Relationship between violence and territorial control.} (Top) Barplots show the count and proportion of district-years that have violence occurring in the 30-day period before peak NDVI, for district-years with and without Taliban presence. (Bottom) Barplots show the count and proportion of district-years that are have Taliban presence, for district-years with and without violence in the 30-day period before peak NDVI. Both figures illustrate a positive relationship between violence and Taliban presence. Our regressions hence estimate their effects jointly.}
\end{figure}

\begin{figure}[hbtp!]
    \centering
	\includegraphics[width=.48\linewidth,trim={0 0 0 0cm},clip,page=3]{1-10-24figA2.pdf}
	\includegraphics[width=.48\linewidth,trim={0 0 0 0cm},clip,page=1]{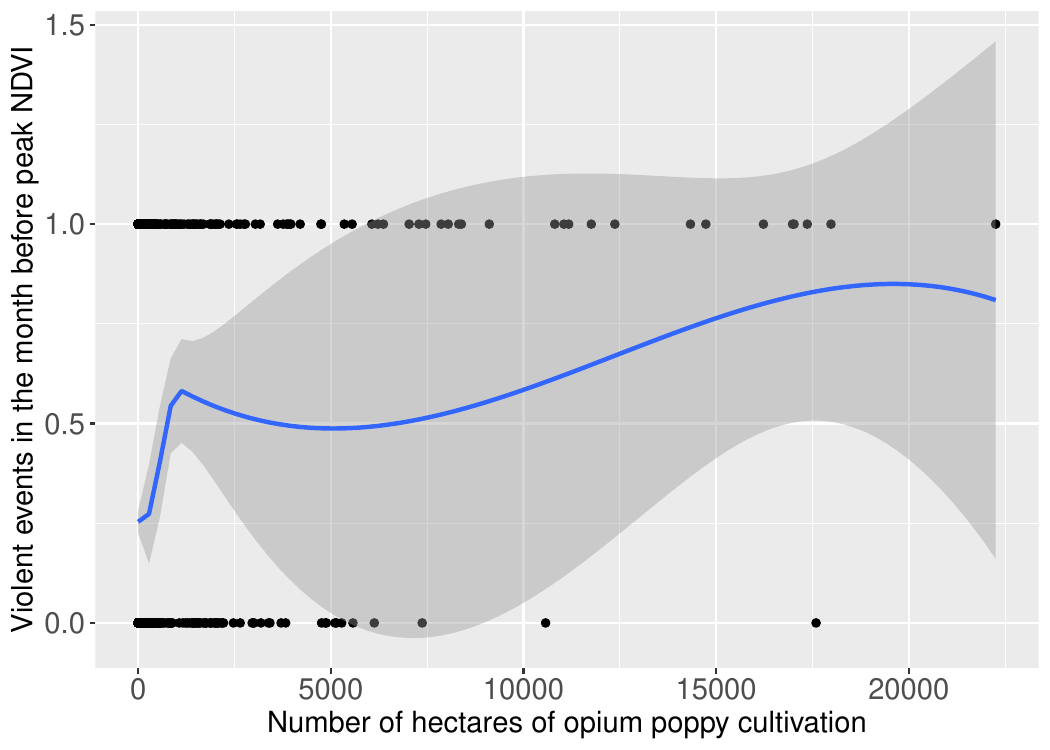}
	\includegraphics[width=.56\linewidth,trim={0 0 0 0cm},clip,page=2]{1-10-24figA2.pdf}
	% \caption*{\footnotesize \textit{Notes:}}
	\label{fig:destinationViolence}
	\caption{\textbf{Relationship between the amount of opium poppy cultivation and increase in harvest in-migration, the occurrence of violence in the month preceding peak NDVI, and Taliban presence at the destination.} Points represent district-years. The blue line is the fit from a local linear (loess) regression, and the gray bands are 95\% confidence intervals. We do not see larger harvest in-migration rates for the largest amounts of cultivation (top-left panel) and our main results treat amount of cultivation as a categorical variable. As a robustness check, we remove districts with over 5000 hectares of cultivation, and find consistent results. Districts with violent events tend to have larger amounts of cultivation; to check that results involving conflict at the destination are not driven by this distribution of cultivation amounts within the ``high'' category, we remove districts with over 5000 hectares of cultivation; results are unaffected (\textit{Figure S4}).}
\end{figure}

\begin{figure}[hbtp!]
    \centering
 	% \hspace*{-8cm}\textbf{C}
	\includegraphics[width=.9\linewidth,trim={0 0 0 0cm},clip]{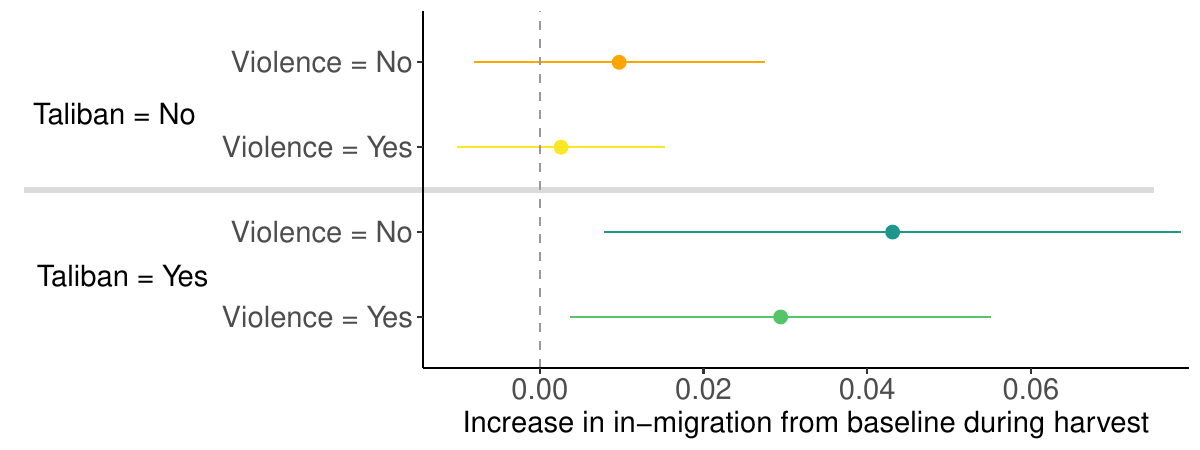}
	% \caption*{\footnotesize \textit{Notes:}}
	\label{fig:destinationRobust1}
      \caption{\textbf{Effect of conflict at the destination on the relationship between cultivation and increase in in-migration during harvest, removing district-years that have over 5000 ha (top 3\%) of opium poppy cultivation.} Regression coefficients estimate the increase in harvest in-migration for high-cultivation destinations with different levels of conflict, relative to districts with no poppy cultivation, no violence and that do not have Taliban presence. Model includes province and year fixed effects, district-year and district controls. Standard errors are clustered at a district level. Bars are 95\% confidence intervals.} 
\end{figure}

\begin{figure}[hbtp!]
    \centering
      \vspace*{-2cm}
	\includegraphics[width=.49\linewidth,trim={0 0 0 0cm},clip,page=1]{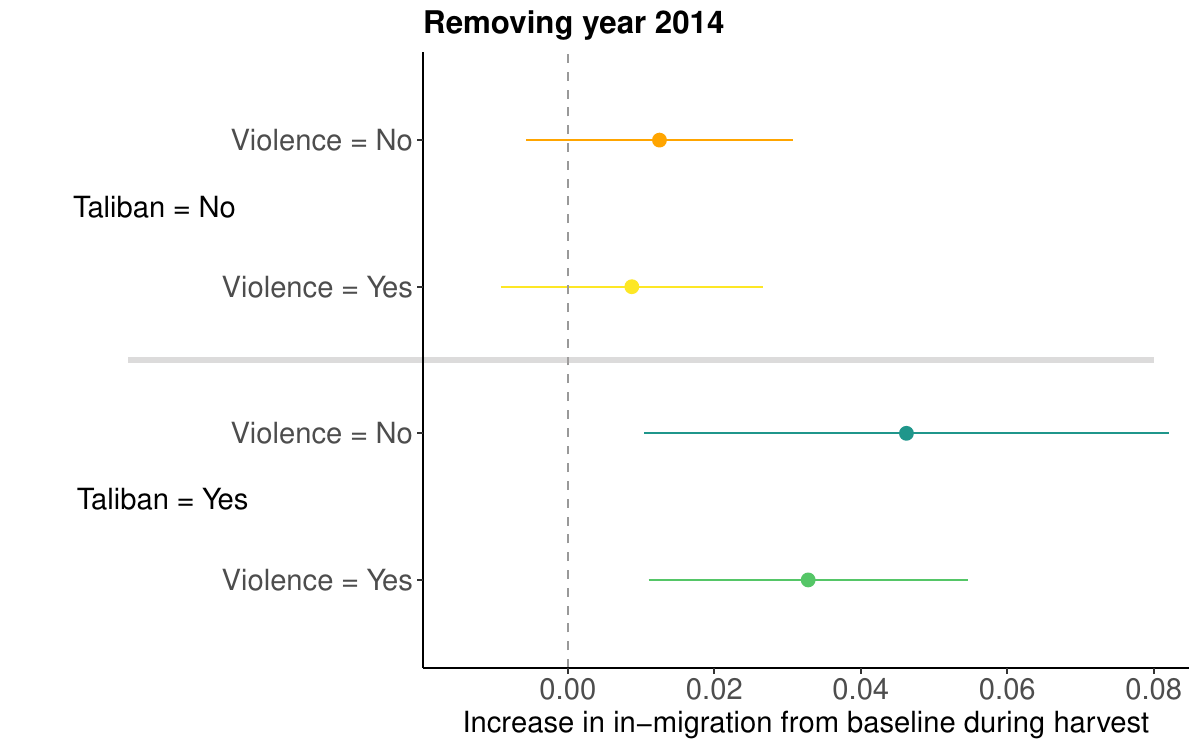}
	\includegraphics[width=.49\linewidth,trim={0 0 0 0cm},clip,page=2]{1-10-24fig3b_robust2.pdf}
	\includegraphics[width=.49\linewidth,trim={0 0 0 0cm},clip,page=3]{1-10-24fig3b_robust2.pdf}
	\includegraphics[width=.49\linewidth,trim={0 0 0 0cm},clip,page=4]{1-10-24fig3b_robust2.pdf}
	\includegraphics[width=.49\linewidth,trim={0 0 0 0cm},clip,page=5]{1-10-24fig3b_robust2.pdf}
	\includegraphics[width=.49\linewidth,trim={0 0 0 0cm},clip,page=6]{1-10-24fig3b_robust2.pdf}
  	% \hspace*{-8cm}\textbf{A}\newline
 \includegraphics[width=.49\linewidth,trim={0 0 0 0cm},clip]{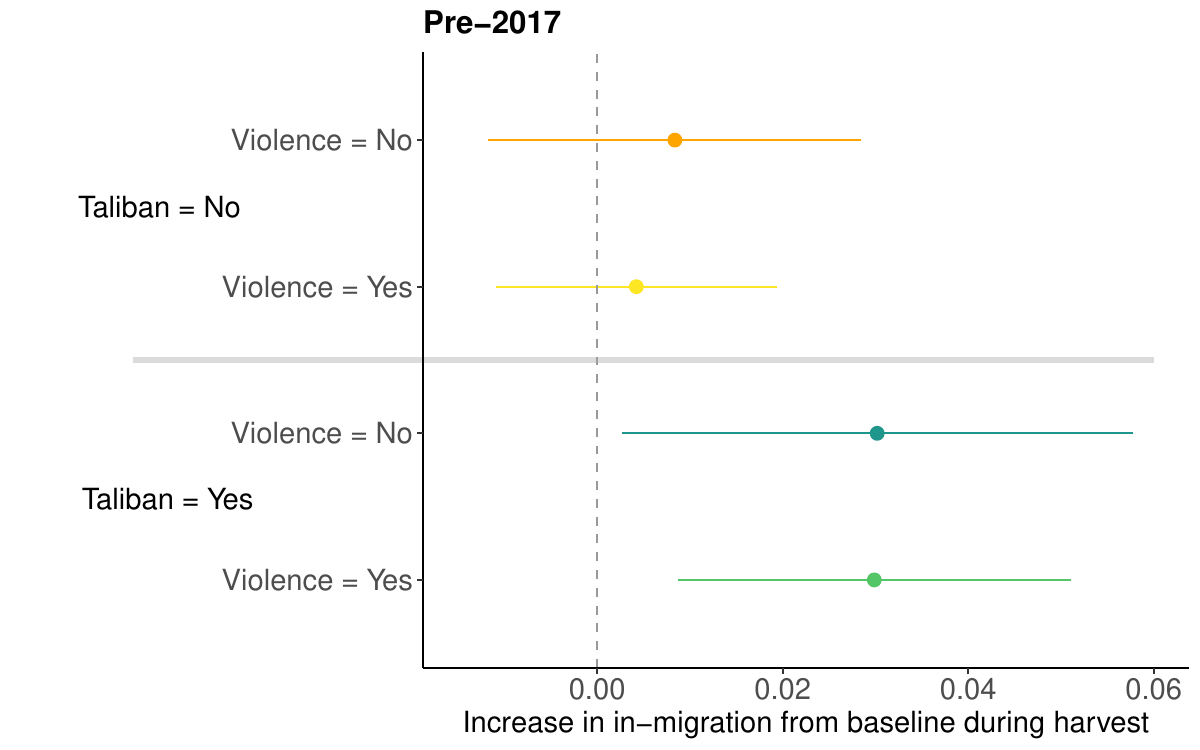}%\newline
 	% \hspace*{-8cm}\textbf{B}
  % \newline
	\includegraphics[width=.49\linewidth,trim={0 0 0 0cm},clip]{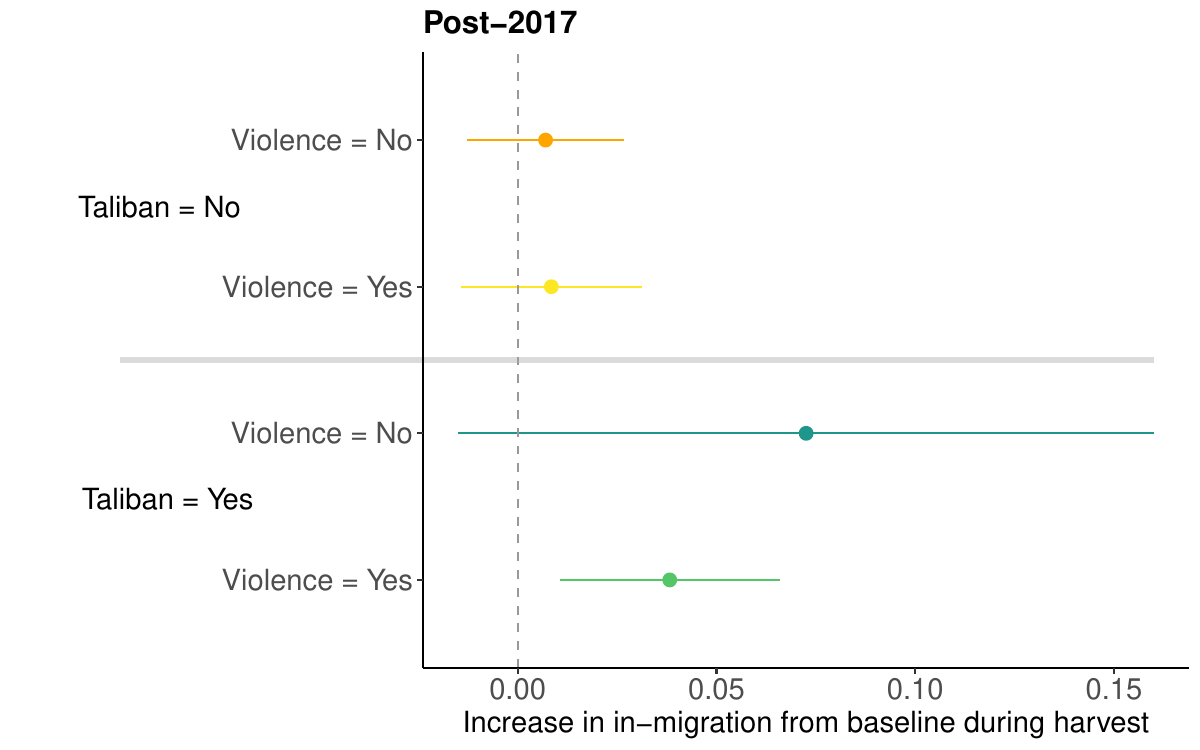}%\newline

	% \caption*{\footnotesize \textit{Notes:}}
	\label{fig:destinationRobust2}
      \caption{\textbf{Effect of conflict at the destination on the relationship between cultivation and increase in in-migration during harvest, removing individual years, and only using pre-2017 and post-2017 years.} Regression coefficients estimate the increase in harvest in-migration for high-cultivation destinations with different levels of conflict, relative to districts with no poppy cultivation, no violence and that do not have Taliban presence. All models include province and year fixed effects, district-year and district controls. Standard errors are clustered at a district level. Bars are 95\% confidence intervals.} 
\end{figure}

\begin{figure}[hbtp!]
    \centering
	\includegraphics[width=.9\linewidth,trim={0 0 0 0cm},clip,page=2]{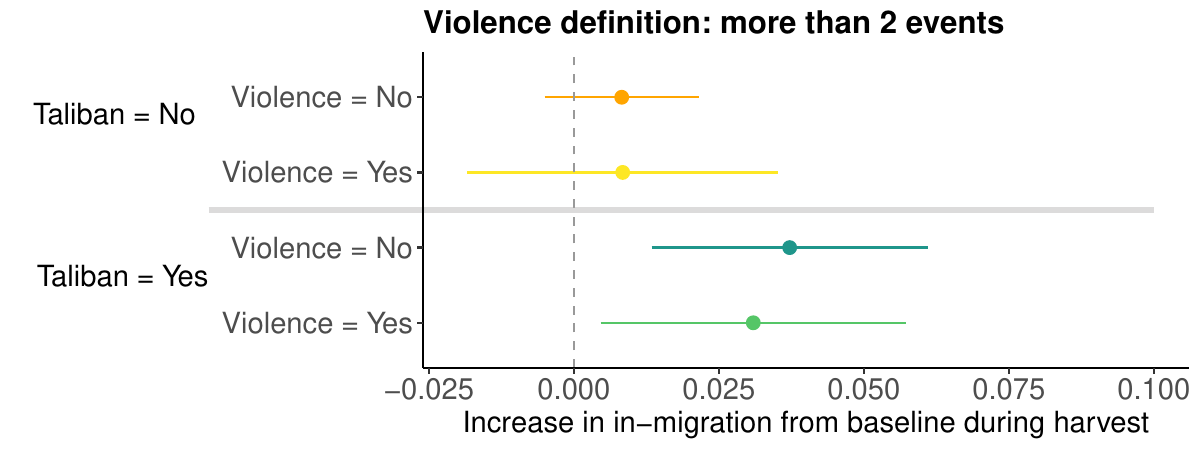}
	\includegraphics[width=.9\linewidth,trim={0 0 0 0cm},clip,page=3]{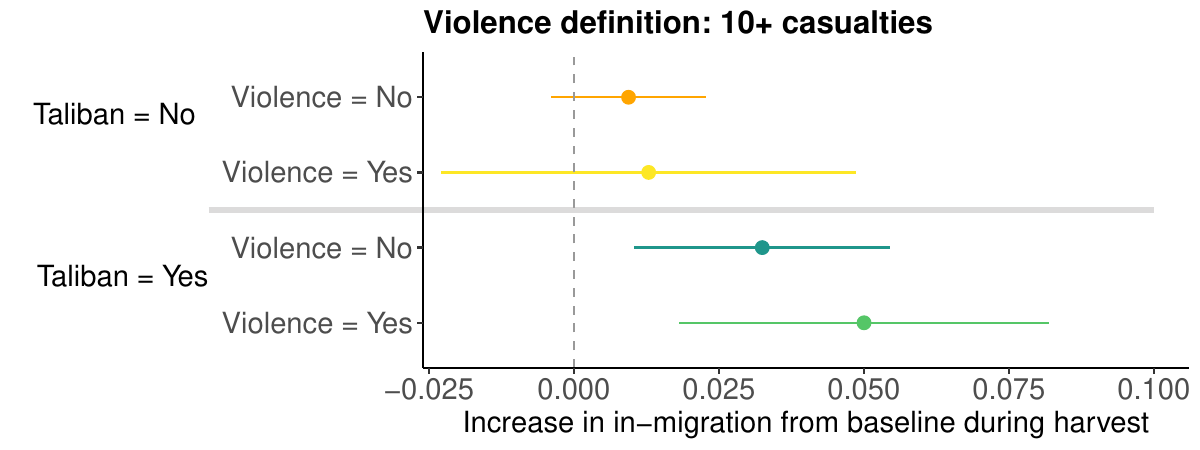}
	\includegraphics[width=.9\linewidth,trim={0 0 0 0cm},clip,page=4]{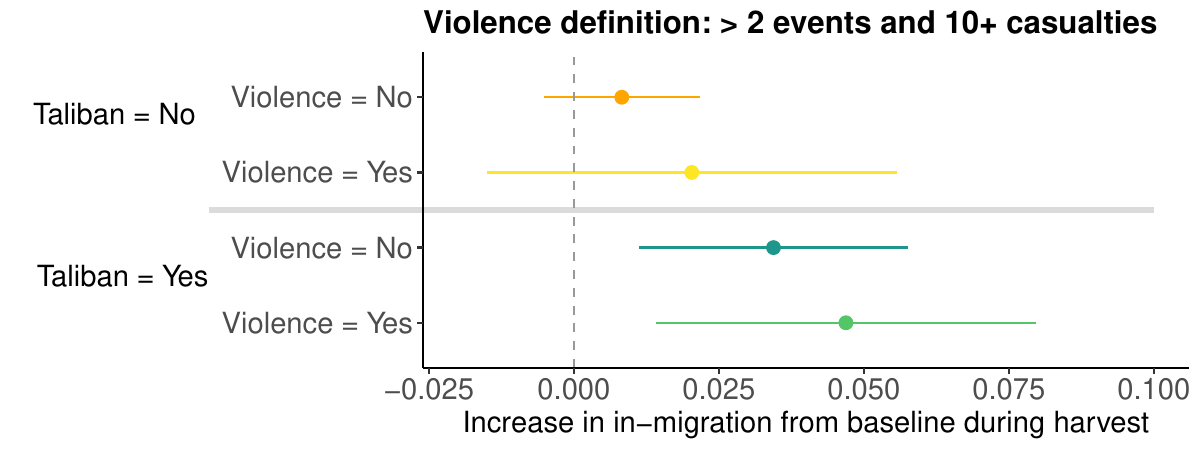}
  	% \hspace*{-8cm}\textbf{A}\newline
	% \caption*{\footnotesize \textit{Notes:}}
	\label{fig:destinationRobust3}
      \caption{\textbf{Effect of conflict at the destination on the relationship between cultivation and increase in in-migration during harvest, using different definitions of violence.} Regression coefficients estimate the increase in harvest in-migration for high-cultivation destinations with different levels of conflict, relative to districts with no poppy cultivation, no violence and that do not have Taliban presence. All models include province and year fixed effects, district-year and district controls. Standard errors are clustered at a district level. Bars are 95\% confidence intervals.} 
\end{figure}

\begin{figure}[hbtp!]
    \centering
	\includegraphics[width=.49\linewidth,trim={0 0 0 0cm},clip,page=1]{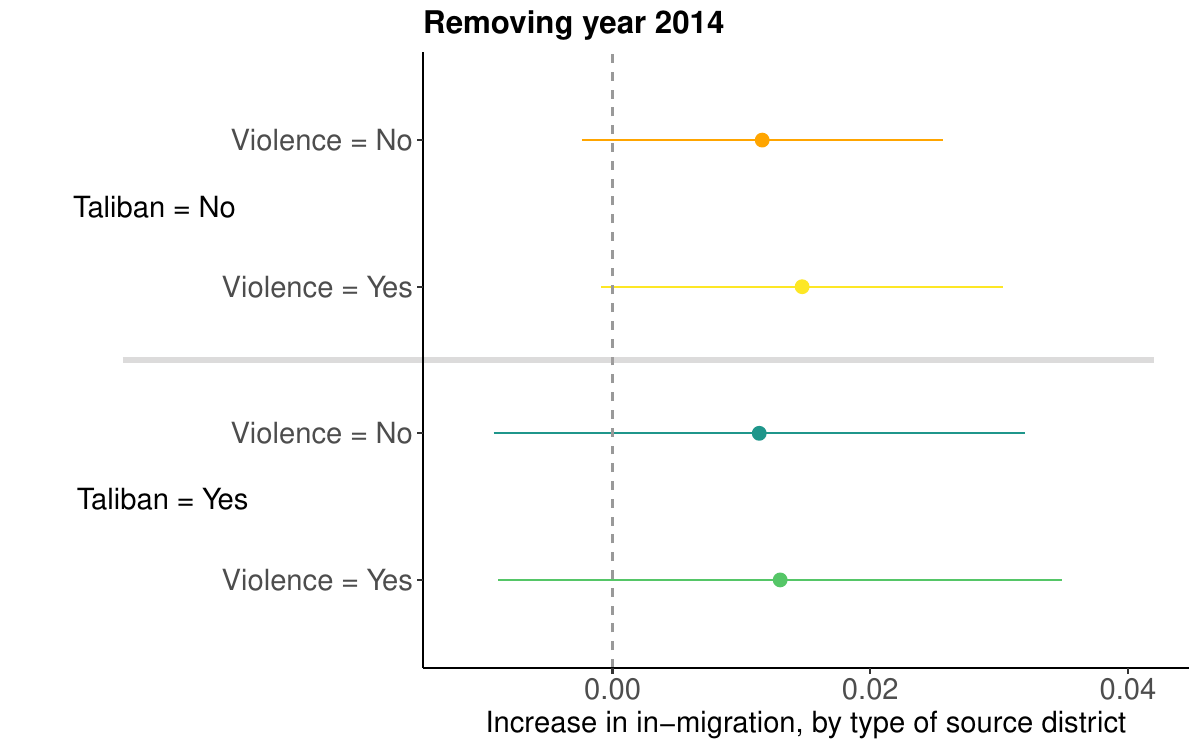}
	\includegraphics[width=.49\linewidth,trim={0 0 0 0cm},clip,page=2]{1-10-24fig4c_robust.pdf}
	\includegraphics[width=.49\linewidth,trim={0 0 0 0cm},clip,page=3]{1-10-24fig4c_robust.pdf}
	\includegraphics[width=.49\linewidth,trim={0 0 0 0cm},clip,page=4]{1-10-24fig4c_robust.pdf}
	\includegraphics[width=.49\linewidth,trim={0 0 0 0cm},clip,page=5]{1-10-24fig4c_robust.pdf}
	\includegraphics[width=.49\linewidth,trim={0 0 0 0cm},clip,page=6]{1-10-24fig4c_robust.pdf}
	% \caption*{\footnotesize \textit{Notes:}}
	\label{fig:sourceViolence}
      \caption{\textbf{Effect of conflict at the source on the relationship between cultivation and increase in in-migration during harvest, removing individual years.} Regression coefficients estimate the increase in harvest proportion of in-migrants to high-cultivation destinations (relative to destinations with no poppy cultivation) from source districts with various characteristics. All models include province and year fixed effects, district-year and district controls. Standard errors are clustered at a district level. Bars are 95\% confidence intervals.} 
\end{figure}

\clearpage
\section{Supplementary Tables}
\begin{table}[!htbp] \centering 
 \scriptsize 
	\def\arraystretch{1.1} % .5
	% \hspace*{-.3cm}
 \setlength\tabcolsep{0pt}
 \begin{tabular*}{\linewidth}{@{\extracolsep{\fill}} lccccccc }\\[-1.8ex]\hline 
\hline \\[-1.8ex] 
 % & \multicolumn{7}{c}{\textit{Dependent variable:}} \\ 
 % \cline{2-8} 
\\[-1.8ex] & (1) & (2) & (3) & (4) & (5) & (6) & (7)\\ 
\\[-1.8ex] & Main  &  Outcome:    & Outcome:   & Outcome:    & Outcome:  & Poppy & Continuous  \\
 & specification & 15-day & 45-day & Longer  & Longer  & cultivation & poppy \\  
 &   & reference &   reference &  window &  continuous & $<$5000 ha & \\  
 &  & period  &  period  &   &    &  & \\  
\\[-1.8ex] 
\hline \\[-1.8ex] 
 Poppy High ($>=$1000ha) & 0.0271$^{***}$ & 0.0271$^{***}$ & 0.0273$^{***}$ & 0.0282$^{***}$ & 0.0263$^{***}$ & 0.0244$^{***}$ &  \\ 
  & (0.0084) & (0.0060) & (0.0087) & (0.0090) & (0.0081) & (0.0091) &  \\ 
  & & & & & & & \\ 
 Poppy Low (1-999) & $-$0.0015 & 0.0007 & $-$0.0014 & $-$0.0006 & $-$0.0007 & $-$0.0011 &  \\ 
  & (0.0028) & (0.0019) & (0.0029) & (0.0030) & (0.0027) & (0.0028) &  \\ 
  & & & & & & & \\ 
 Continuous poppy (ha, log) &  &  &  &  &  &  & 0.0020$^{**}$ \\ 
  &  &  &  &  &  &  & (0.0008) \\ 
  & & & & & & & \\ 
\hline \\[-1.8ex] 
Observations & 1,405 & 1,372 & 1,422 & 1,405 & 1,404 & 1,371 & 1,405 \\ 
R$^{2}$ & 0.1313 & 0.1823 & 0.1293 & 0.1440 & 0.1414 & 0.0998 & 0.1037 \\ 
Adjusted R$^{2}$ & 0.0939 & 0.1462 & 0.0922 & 0.1071 & 0.1044 & 0.0600 & 0.0657 \\ 
\hline 
\hline \\[-1.8ex] 
\end{tabular*} 
\caption{\textbf{Robustness checks for the overall effect of cultivation on the increase in in-migration during the harvest.} Column (1) reproduces the main results shown in Figure 2D. Columns (2)-(5) vary parameters used to measure migration: (2) uses a 15-day reference period to define in-migration and (3) uses a 45-day period; (4) uses a 1-45 day window to determine maximum harvest in-migration, and (5) additionally takes the mean over a 14-day period (instead of 7). Column (6) removes district-years with the highest cultivation of 5000 ha or more of opium poppy (top 3\%). %This checks that our results are not driven by the distribution of amounts of cultivation within the ``high-cultivation'' category. 
Column (7) uses a continuous variable for poppy cultivation. All models include province and year fixed effects, district-year and district controls. Standard errors are clustered at a district level.  $^{*}$p$<$0.1; $^{**}$p$<$0.05; $^{***}$p$<$0.01.} 
  \label{tab:regResultsOverall} 
\end{table} 

% Table created by stargazer v.5.2.3 by Marek Hlavac, Social Policy Institute. E-mail: marek.hlavac at gmail.com
% Date and time: Fri, Dec 20, 2024 - 02:40:18 PM
\begin{table}[!htbp] \centering 
 \scriptsize 
	\def\arraystretch{1.1} % .5
	% \hspace*{-.3cm}
 \setlength\tabcolsep{0pt}
 \begin{tabular*}{.9\linewidth}{@{\extracolsep{\fill}} lcccc }\\[-1.8ex]\hline 
\hline \\[-1.8ex] 
 % & \multicolumn{7}{c}{\textit{Dependent variable:}} \\ 
 % \cline{2-8} 
\\[-1.8ex] & (1) & (2) & (3) & (4) \\ 
\\[-1.8ex] & Subtract 14 days & Add 14 days &  Districts with majority & Districts with  minority   \\
 &   & &same  peak NDVI & same  peak NDVI  \\  
\\[-1.8ex] 
\hline \\[-1.8ex] 
 Poppy High ($>=$1000ha) & 0.0151$^{**}$ & 0.0230$^{***}$ & 0.0317$^{**}$ & 0.0212$^{***}$ \\ 
  & (0.0068) & (0.0056) & (0.0133) & (0.0062) \\ 
  & & & & \\ 
 Poppy Low (1-999) & $-$0.0034 & 0.0004 & $-$0.0022 & $-$0.0012 \\ 
  & (0.0024) & (0.0022) & (0.0042) & (0.0035) \\ 
  & & & & \\ 
\hline \\[-1.8ex] 
Observations & 1,405 & 1,405 & 696 & 709 \\ 
R$^{2}$ & 0.1161 & 0.1392 & 0.1933 & 0.1568 \\ 
Adjusted R$^{2}$ & 0.0781 & 0.1021 & 0.1212 & 0.0830 \\ 
\hline 
\hline \\[-1.8ex] 
\end{tabular*} 
\caption{\textbf{Additional robustness checks for NDVI-based measure.} Same as Table S1, but for robustness to errors relating to the satellite-based measurement of the peak NDVI period. Column (1) perturbs the peak NDVI dates by subtracting 14 days to the estimated dates for every district-year, while (2) displays results adding 14 days. Column (3) only includes districts in which the modal date occurs among majority of the pixels, i.e., more than 50\% of pixels in the district have the same 16-day period corresponding to peak NDVI. These districts can be thought of as being relatively uniform, where peak agricultural activity, opium or otherwise, is measured more precisely. Column (4) includes districts in which the opposite is true. All models include province and year fixed effects, district-year and district controls. Standard errors are clustered at a district level.  $^{*}$p$<$0.1; $^{**}$p$<$0.05; $^{***}$p$<$0.01.} 
  \label{tab:robust2} 
\end{table} 

\begin{table}[!htbp] \centering 
\footnotesize
	\def\arraystretch{1.1} % .5
	% \hspace*{-1.8cm}
\begin{tabular}{@{\extracolsep{5pt}}lccc} 
\\[-1.8ex]\hline 
\hline \\[-1.8ex] 
 & \multicolumn{3}{c}{\textit{Dependent variable:}} \\ 
\cline{2-4} 
\\[-1.8ex] & (1) & (2) & (3)\\ 
& Same district  & Previous district & Previous district\\ 
& (day 30) & (day 30) & (90-day)\\ 
\hline \\[-1.8ex] 
 Poppy High ($\geq$1000ha) & $-$0.0703$^{***}$ & 0.0694$^{***}$ & 0.0670$^{***}$ \\ 
  & (0.0193) & (0.0167) & (0.0183) \\ 
  & & & \\ 
 Poppy Low (1-999) & $-$0.0092 & 0.0157$^{**}$ & 0.0158$^{*}$ \\ 
  & (0.0089) & (0.0079) & (0.0090) \\ 
  & & & \\ 
\hline \\[-1.8ex] 
Observations & 1,342 & 1,257 & 1,354 \\ 
R$^{2}$ & 0.1632 & 0.2374 & 0.2791 \\ 
Adjusted R$^{2}$ & 0.1254 & 0.2005 & 0.2468 \\ 
\hline 
\hline \\[-1.8ex] 
% \textit{Note:}  & \multicolumn{3}{r}{$^{*}$p$<$0.1; $^{**}$p$<$0.05; $^{***}$p$<$0.01} \\ 
\end{tabular} 
  \caption{\textbf{Return of seasonal migrants.} Regressions use different outcome variables that take into account subsequent movement: column (1) displays the increase in proportion of in-migrants during the harvest that are observed to still be in the same district 30 days after being observed, i.e., day $t+30$, and column (2) the increase in proportion of in-migrants during the harvest that are observed to have returned to their previous district (i.e., their observed district on day $t-30$), on day $t+30$. Column (3) shows the increase in proportion of in-migrants during the harvest that are observed to have returned to their previous district (i.e., their observed district on day $t-30$), any time in the 90-day window, days $t+1$ to $t+90$. All models include province and year fixed effects, district-year and district controls. Standard errors are clustered at a district level. $^{*}$p$<$0.1; $^{**}$p$<$0.05; $^{***}$p$<$0.01.} 
  \label{tab:morerobust} 
\end{table}

\clearpage
\section{Effects of violence along routes of entry}

Our main results describe the manner in which violence at the source and destination locations shape seasonal migration flows. Here, we examine whether violent events that occur along transportation routes influence patterns of seasonal migration. More specifically, for each district, we consider violent events which occur along key routes of entry to that district.

\paragraph{Data}
\begin{itemize}
    \item Road networks: We rely on geospatial data describing the network of major roads in Afghanistan in 2016.\footnote{Collected by OpenStreetMap and Geofabrik GmbH, available at \url{https://download.geofabrik.de/asia/afghanistan.html}.} For the purposes of this exercise, we hold the road network fixed over time. 
    \item Violent events: Data on violent events are obtained from the Uppsala Conflict Data Program \cite{Sundberg:2013aa} (UCDP Georeferenced Event Dataset (GED) Global version 21.1),\footnote{\url{https://ucdp.uu.se/downloads/}} processed as described in \textit{Materials and Methods}. 
\end{itemize}

\paragraph{Road violence}

\begin{figure}
    \centering
    \includegraphics[width=\linewidth]{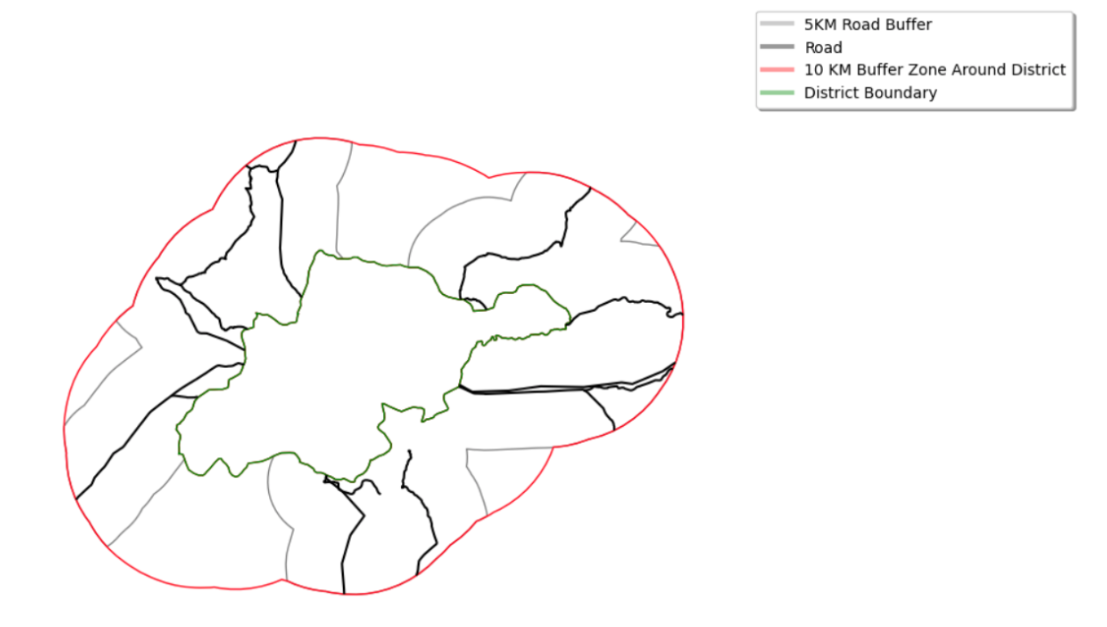}
    \caption{\textbf{Identifying routes of entry for each district.} The green boundary is the district boundary for an example district. We consider roads within a 10 km buffer zone around the district (red boundary). To measure road violence for the district, we consider violent events occurring within 5 km of these roads during the month prior to the peak harvest date of the district.}
    \label{fig:road-dist}
\end{figure}

For each district \textit{i}, we identify all sections of road leading to the district \textit{i} within 10 km of the district boundary. For each road in this region, we identify all violent events that occur in the month prior to the peak harvest date (similar to \cite{alfano_spatial_2022} we consider only those violent events which are 5 km or closer to each road). Figure~\ref{fig:road-dist} illustrates this strategy for a single district. Once we identify this set of violent events, we define a binary variable, which equals 1 if we observe any violent events occurring along the entry routes to that district. 

We note that the accuracy of this measure of road violence is limited by the geographical precision of violent events recorded in our data. The locations of violent events that we consider are known to either have occurred in a district (48\% of events), within a 25 km radius around a known point (19\% of events), or at an exact location (33\% of events); this imprecision might introduce measurement errors in the road violence variable.

\paragraph{Estimation}

We then run a regression similar to the specification defined in \textit{Materials and Methods} (Equation 4). Here, $V$ is the indicator for violent events occurring within the district, in the month preceding the peak harvest, while $RV$ indicates violence along entry routes as defined above:
\begin{equation}
\label{eq:conflictDestRoadViolence}
     \begin{aligned}
E_{dy} & = \beta_{hvt} H_{dy}*V_{dy}*T_{d} + \beta_{lvt} L_{dy}*V_{dy}*T_{d} \\
 &+ \beta_{hv} H_{dy}*V_{dy}  + \beta_{lv} L_{dy}*V_{dy} \\
 &+ \beta_{hrt} H_{dy}*RV_{dy}*T_{d} + \beta_{lrt} L_{dy}*RV_{dy}*T_{d} \\
 &+ \beta_{hr} H_{dy}*RV_{dy}  + \beta_{lr} L_{dy}*RV_{dy} \\
 &+ \beta_{ht} H_{dy}*T_{d} + \beta_{lt} L_{dy}*T_{d} \\
 & + \beta_{vt} V_{dy}*T_{d}\\
  & + \beta_{rt} RV_{dy}*T_{d}\\
&+ \beta_{h} H_{dy} + \beta_{l} L_{dy} +    \beta_{v} V_{dy}  +  \beta_{r} RV_{dy}  +    \beta_{t} T_{d}\\
    & +  \boldsymbol{\beta} \mathbf{X_{dy}} + \gamma_p  + \lambda_y +  \epsilon_{dy}
\end{aligned}
\end{equation}

\begin{figure}[hbtp!]
    \centering
	\includegraphics[width=.8\linewidth,trim={0 0 0 0cm},clip,page=1]{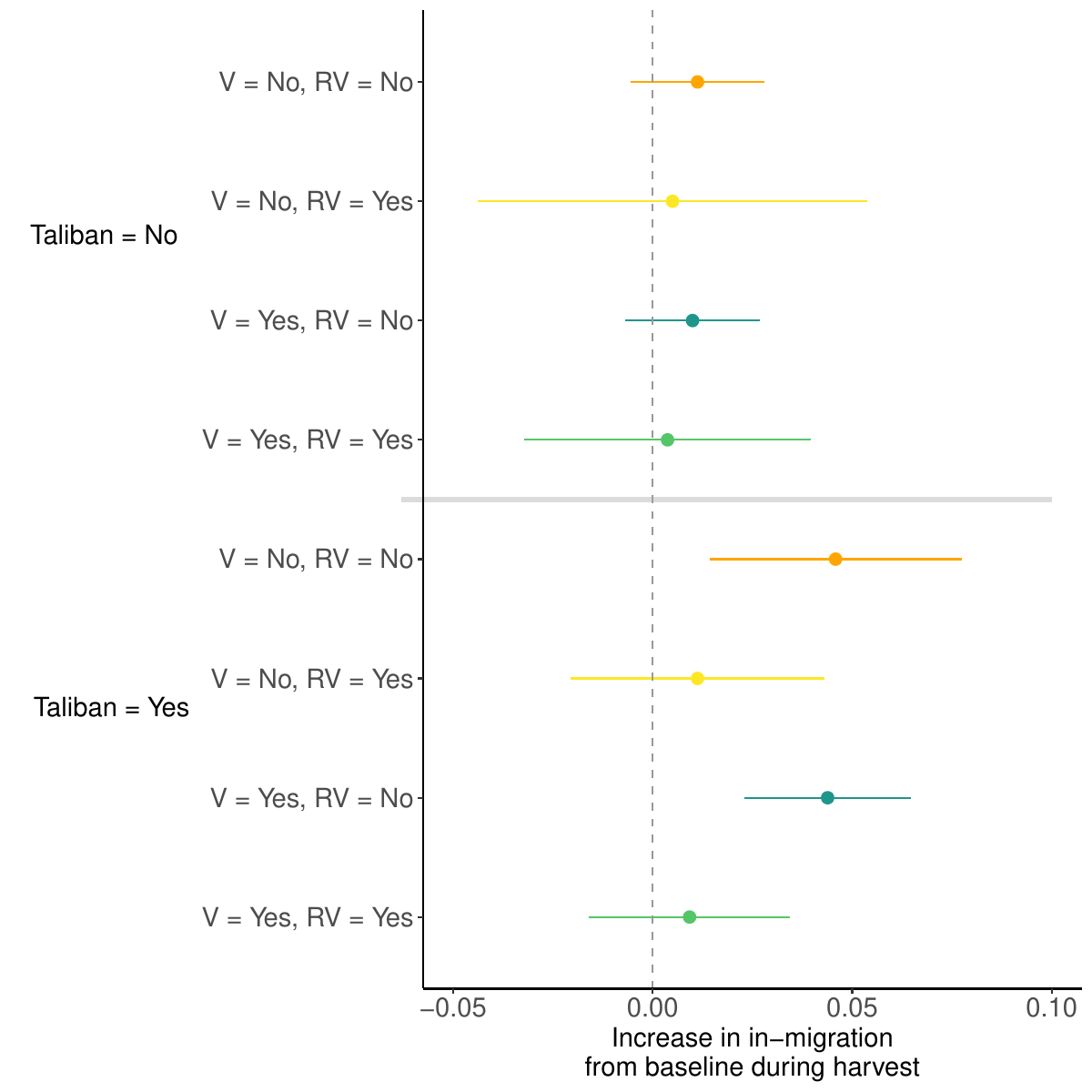}
	\caption{\textbf{Effect of violence along routes of entry in high-cultivation districts.} Regression coefficients estimate the increase in harvest in-migration, relative to districts with no poppy cultivation, no violence, no violence along entry routes and no Taliban presence, controlling for district-level characteristics, unobserved province-level characteristics, and common effects over time. V stands for violence and RV stands for road violence. Bars indicate 95\% confidence intervals.}
    \label{fig:road-violence}
\end{figure}

The coefficients reported in Figure \ref{fig:road-violence} represent, for high-cultivation districts, the estimated difference in the outcome (increase in harvest in-migration) for the groups of interest compared to a reference group with no poppy cultivation, no violence, no violence along entry routes and no Taliban presence. The estimate is obtained by summing the relevant regression coefficients (see \textit{Materials and Methods}).

\paragraph{Conclusion}
The findings on violence at the destination are similar to Figure 3B, with no significant evidence suggesting that this type of violence had an effect on in-migration. However, for destinations with Taliban presence, the there is less in-migration when road violence is present, both when violence in the destination is present and absent (estimate of the difference is -0.035, 95\% C.I.: (-0.057, -0.012), $P = 0.002$). This suggests that violence on road networks bordering a district with high poppy cultivation and Taliban presence reduces in-migration. %This supports our finding that the potential occurrence of violence close to an individual's destination does not serve as a deterrent; instead the actual disruption of travel along major roads could hinder migrants from arriving at their planned destination. 

\clearpage
\section{Descriptive Statistics and Full Regression Results}

\begin{figure}[hbtp!] 
    \hspace*{-.7cm}
    \centering
	\includegraphics[width=1.15\linewidth,trim={0 0 0 .6cm},clip,page=1]{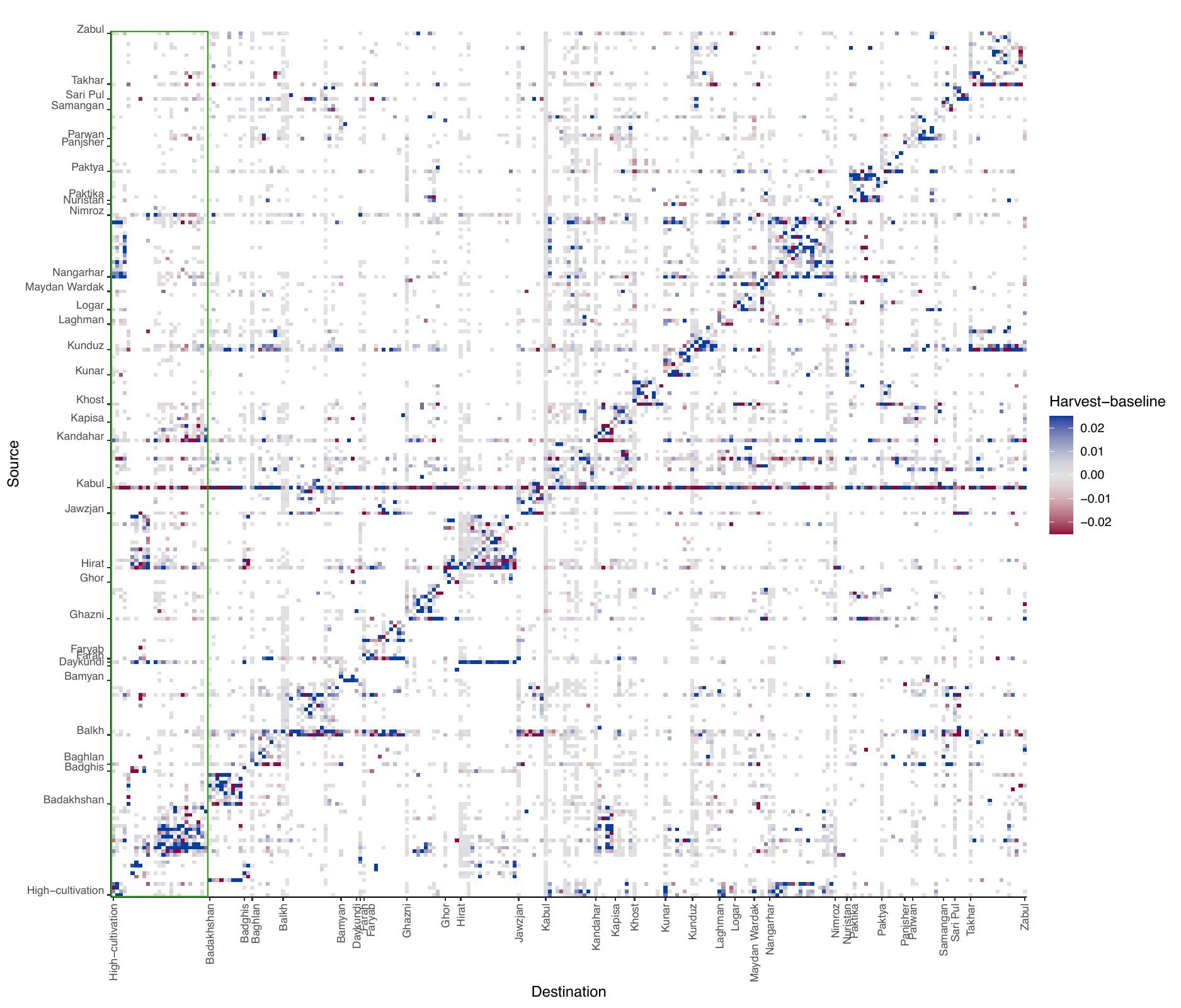}
	\caption{\textbf{Origin-destination matrix of seasonal migration for 2015.} Each row represents a source district, each column represents a destination district. Districts are arranged so that high-cultivation districts are displayed first; remaining districts are grouped by province and ordered alphabetically. Cell colors represent the difference between harvest and baseline daily average proportions of in-migrants to each destination originating from each of the source districts, during the harvest compared to the baseline. These are calculated in the same way as outcome variables involving the composition of in-migrants (see \textit{Measuring harvest migration}). Values are top and bottom coded at $\pm 2.5\%$. Green rectangle represents movements into high-cultivation destinations.}
    \label{fig:ODmatrix}
\end{figure}

% Table created by stargazer v.5.2.3 by Marek Hlavac, Social Policy Institute. E-mail: marek.hlavac at gmail.com
% Date and time: Wed, Aug 14, 2024 - 10:27:16 AM
\begin{table}[!htbp] \centering 
  \label{asdf} 
\begin{tabular}{@{\extracolsep{5pt}}lccccc} 
\\[-1.8ex]\hline 
\hline \\[-1.8ex] 
Statistic & \multicolumn{1}{c}{N} & \multicolumn{1}{c}{Mean} & \multicolumn{1}{c}{St. Dev.} & \multicolumn{1}{c}{Min} & \multicolumn{1}{c}{Max} \\ 
\hline \\[-1.8ex] 
Land area (km$^2$) & 1,420 & 1,333.10 & 1,793.30 & 25.83 & 21,864.30 \\ 
Other cultivation (ha) & 1,420 & 8,521.51 & 10,658.49 & 1.00 & 70,862.50 \\ 
Population & 1,420 & 89,218.27 & 243,207.10 & 5,640 & 3,678,034 \\ 
Provincial capital & 1,420 & 0.14 & 0.35 & 0 & 1 \\ 
Fraction living in cultivated areas & 1,420 & 0.08 & 0.06 & 0.001 & 0.30 \\ 
Road density (km/km$^2$) & 1,420 & 1.20 & 1.36 & 0.00 & 12.18 \\ 
Fraction living in built-up areas & 1,420 & 0.86 & 0.10 & 0.47 & 1.00 \\ 
Fraction of roads that are tracks and paths & 1,414 & 0.89 & 0.13 & 0.01 & 1.00 \\ 
Health facilities per 100,000 & 1,420 & 7.84 & 4.76 & 0.00 & 34.65 \\ 
Ethnic diversity & 1,411 & 0.21 & 0.21 & 0.00 & 0.76 \\ 
Fraction of barren land area & 1,420 & 0.73 & 0.22 & 0.01 & 1.00 \\ 
\hline \\[-1.8ex] 
\end{tabular} 
\caption{\textbf{Descriptive statistics for covariates in the regression sample.}}
\end{table}

% Table created by stargazer v.5.2.3 by Marek Hlavac, Social Policy Institute. E-mail: marek.hlavac at gmail.com
% Date and time: Wed, Aug 14, 2024 - 11:26:41 AM
\begin{table}[!htbp] \centering 
  \label{qwer} 
\begin{tabular}{lccccc} 
\\[-1.8ex]\hline 
\hline \\[-1.8ex] 
Statistic & \multicolumn{1}{c}{N} & \multicolumn{1}{c}{Mean} & \multicolumn{1}{c}{St. Dev.} & \multicolumn{1}{c}{Min} & \multicolumn{1}{c}{Max} \\ 
\hline \\[-1.8ex] 
Poppy cultivation (ha) & 1,420 & 430.18 & 1,811.52 & 0 & 22,256 \\ 
High poppy cultivation & 1,420 & 0.09 & 0.28 & 0 & 1 \\ 
Low poppy cultivation & 1,420 & 0.25 & 0.43 & 0 & 1 \\ 
Violent events & 1,420 & 0.30 & 0.46 & 0 & 1 \\ 
Taliban presence & 1,420 & 0.37 & 0.48 & 0 & 1 \\ 
Outcome: Increase in in-migration & 1,420 & 0.01 & 0.03 & $-$0.12 & 0.42 \\ 
Outcome: Increase in proportion from &  & &  &  &  \\ 
\hspace{.5cm}high cultivation districts & 1,419 & 0.03 & 0.07 & $-$0.68 & 0.53 \\ 
\hspace{.5cm}low cultivation districts & 1,404 & 0.03 & 0.07 & $-$0.44 & 0.49 \\ 
\hspace{.5cm}high, violent, Taliban districts & 1,420 & 0.02 & 0.05 & $-$0.28 & 0.56 \\ 
\hspace{.5cm}high, violent, non-Taliban districts & 1,419 & 0.01 & 0.04 & $-$0.39 & 0.34 \\ 
\hspace{.5cm}high, non-violent, Taliban districts & 1,420 & 0.01 & 0.04 & $-$0.30 & 0.44 \\ 
\hspace{.5cm}high, non-violent, non-Taliban districts & 1,420 & 0.01 & 0.03 & $-$0.36 & 0.28 \\ 
\hline \\[-1.8ex] 
\end{tabular} 
  \caption{\textbf{Descriptive statistics for variables of interest in the regression sample.}} 
\end{table}

\clearpage

% Table created by stargazer v.5.2.3 by Marek Hlavac, Social Policy Institute. E-mail: marek.hlavac at gmail.com
% Date and time: Wed, Aug 14, 2024 - 04:52:11 PM
\begin{table}[!htbp] \centering 
  % \caption{} 
  \label{1234} 
\scriptsize 
\setlength{\tabcolsep}{2pt} % Default value: 6pt
\begin{tabular}{@{\extracolsep{0pt}}lccccccccc} 
\\[-1.8ex]\hline 
\hline \\[-1.8ex] 
 % & \multicolumn{9}{c}{\textit{Dependent variable:}} \\ 
% \cline{2-10} 
% \\[-1.8ex] & \multicolumn{9}{c}{maxIn} \\ 
% [-1.8ex] \\ 
& Fig 2D & Fig 3B & \multicolumn{2}{c}{Fig 4B} & \multicolumn{4}{c}{Fig 4C} & Fig 5A \\ 
\cmidrule(lr){4-5}                  
    \cmidrule(lr){6-9} & (1) & (2) & (3) & (4) & (5) & (6) & (7) & (8) & (9)\\ 
\hline \\[-1.8ex] 
  Poppy High ($>$1000ha) & 0.0271$^{***}$ & 0.0098 & 0.0473$^{***}$ & 0.0070 & 0.0206$^{**}$ & 0.0153$^{*}$ & 0.0156$^{**}$ & 0.0111$^{*}$ & 0.0261$^{***}$ \\ 
  & (0.0084) & (0.0085) & (0.0154) & (0.0110) & (0.0100) & (0.0089) & (0.0074) & (0.0059) & (0.0090) \\ 
 [-1.8ex] \\
 Poppy Low (1-999) & $-$0.0015 & $-$0.0050$^{*}$ & 0.0205$^{***}$ & 0.0082 & 0.0053 & 0.0095$^{***}$ & 0.0032 & 0.0038 & 0.0027 \\ 
  & (0.0028) & (0.0027) & (0.0059) & (0.0085) & (0.0042) & (0.0032) & (0.0031) & (0.0026) & (0.0038) \\ 
 [-1.8ex] \\
  Land area (km$^2$, log) & $-$0.0011 & $-$0.0007 & 0.0002 & 0.0071$^{**}$ & $-$0.0010 & 0.0007 & $-$0.0023 & 0.0026$^{**}$ & $-$0.0008 \\ 
  & (0.0017) & (0.0017) & (0.0040) & (0.0032) & (0.0023) & (0.0016) & (0.0024) & (0.0013) & (0.0018) \\ 
 [-1.8ex] \\
  Other cultivation (ha, log) & 0.0007 & 0.0003 & $-$0.0019 & $-$0.0039$^{**}$ & 0.0008 & 0.0004 & $-$0.0013 & $-$0.0001 & 0.0002 \\ 
  & (0.0008) & (0.0008) & (0.0016) & (0.0017) & (0.0013) & (0.0010) & (0.0012) & (0.0007) & (0.0009) \\ 
 [-1.8ex] \\
  Population & $-$0.0000 & $-$0.0000 & $-$0.0000 & $-$0.0000$^{*}$ & 0.0000 & $-$0.0000$^{**}$ & $-$0.0000 & $-$0.0000 & $-$0.0000 \\ 
  & (0.0000) & (0.0000) & (0.0000) & (0.0000) & (0.0000) & (0.0000) & (0.0000) & (0.0000) & (0.0000) \\ 
 [-1.8ex] \\
  Provincial Capital & $-$0.0009 & $-$0.0001 & 0.0043 & 0.0023 & $-$0.0028 & $-$0.0021 & 0.0021 & 0.0005 & 0.0040 \\ 
  & (0.0026) & (0.0025) & (0.0062) & (0.0063) & (0.0040) & (0.0029) & (0.0034) & (0.0023) & (0.0027) \\ 
 [-1.8ex] \\
  Fraction of population & $-$0.0586$^{*}$ & $-$0.0506 & 0.1226 & 0.0985 & 0.1432$^{***}$ & 0.0251 & $-$0.0584 & 0.0331 & 0.0027 \\ 
  in cultivated areas& (0.0334) & (0.0322) & (0.0767) & (0.0666) & (0.0548) & (0.0297) & (0.0397) & (0.0301) & (0.0398) \\ 
 [-1.8ex] \\
  Road density & $-$0.0004 & $-$0.0006 & $-$0.0017 & $-$0.0013 & $-$0.0029$^{*}$ & 0.0018 & 0.0013 & 0.0003 & 0.0005 \\ 
  & (0.0013) & (0.0012) & (0.0031) & (0.0026) & (0.0018) & (0.0013) & (0.0016) & (0.0011) & (0.0014) \\ 
 [-1.8ex] \\
  Fraction of population & $-$0.0490$^{*}$ & $-$0.0384 & 0.1485$^{**}$ & 0.0522 & 0.0757$^{*}$ & 0.0325 & $-$0.0011 & 0.0324 & $-$0.0274 \\ 
  in built-up areas& (0.0282) & (0.0252) & (0.0605) & (0.0604) & (0.0424) & (0.0280) & (0.0301) & (0.0253) & (0.0421) \\ 
 [-1.8ex] \\
  Fraction of roads that & $-$0.0045 & $-$0.0020 & 0.0195 & $-$0.0032 & 0.0121 & $-$0.0174$^{*}$ & 0.0151 & 0.0083 & 0.0142 \\ 
  are tracks and paths& (0.0114) & (0.0116) & (0.0203) & (0.0222) & (0.0136) & (0.0099) & (0.0099) & (0.0115) & (0.0093) \\ 
 [-1.8ex] \\
  Health facilities & $-$0.0002 & $-$0.0002 & $-$0.0005 & $-$0.0006 & 0.0004 & 0.0004 & $-$0.0012$^{***}$ & 0.0005$^{*}$ & $-$0.00004 \\ 
  per 100,000& (0.0003) & (0.0003) & (0.0007) & (0.0005) & (0.0004) & (0.0003) & (0.0004) & (0.0003) & (0.0004) \\ 
 [-1.8ex] \\
  Ethnic diversity & $-$0.0103$^{**}$ & $-$0.0101$^{**}$ & $-$0.0122 & $-$0.0235$^{*}$ & $-$0.0059 & $-$0.0036 & $-$0.0017 & $-$0.0045 & $-$0.0072 \\ 
  & (0.0052) & (0.0051) & (0.0120) & (0.0128) & (0.0063) & (0.0054) & (0.0058) & (0.0050) & (0.0053) \\ 
 [-1.8ex] \\
  Fraction of barren & 0.0033 & $-$0.0001 & 0.0507$^{***}$ & $-$0.0109 & 0.0111 & 0.0049 & 0.0245$^{*}$ & 0.0104$^{*}$ & 0.0047 \\ 
  land area & (0.0092) & (0.0087) & (0.0191) & (0.0152) & (0.0111) & (0.0072) & (0.0126) & (0.0063) & (0.0076) \\ 
 [-1.8ex] \\
  Violent events &  & 0.0025 &  &  &  &  &  &  &  \\ 
  &  & (0.0025) &  &  &  &  &  &  &  \\ 
 [-1.8ex] \\
  Taliban presence &  & 0.0007 &  &  &  &  &  &  &  \\ 
  &  & (0.0028) &  &  &  &  &  &  &  \\ 
 [-1.8ex] \\
  Poppy Low:Violence &  & $-$0.0042 &  &  &  &  &  &  &  \\ 
  &  & (0.0038) &  &  &  &  &  &  &  \\ 
 [-1.8ex] \\
  Poppy High:Violence &  & $-$0.0038 &  &  &  &  &  &  &  \\ 
  &  & (0.0108) &  &  &  &  &  &  &  \\ 
 [-1.8ex] \\
  Poppy Low:Taliban &  & 0.0054 &  &  &  &  &  &  &  \\ 
  &  & (0.0047) &  &  &  &  &  &  &  \\ 
 [-1.8ex] \\
  Poppy High:Taliban &  & 0.0310$^{*}$ &  &  &  &  &  &  &  \\ 
  &  & (0.0166) &  &  &  &  &  &  &  \\ 
 [-1.8ex] \\
  Violence:Taliban &  & $-$0.0072$^{**}$ &  &  &  &  &  &  &  \\ 
  &  & (0.0035) &  &  &  &  &  &  &  \\ 
 [-1.8ex] \\
  Low:Violence:Taliban &  & 0.0140$^{*}$ &  &  &  &  &  &  &  \\ 
  &  & (0.0084) &  &  &  &  &  &  &  \\ 
 [-1.8ex] \\
  High:Violence:Taliban &  & 0.0004 &  &  &  &  &  &  &  \\ 
  &  & (0.0173) &  &  &  &  &  &  &  \\ 
 [-1.8ex] \\
  Eradication percent &  &  &  &  &  &  &  &  & 0.0056 \\ 
  &  &  &  &  &  &  &  &  & (0.0052) \\ 
 [-1.8ex] \\
  Low:Eradication &  &  &  &  &  &  &  &  & $-$0.0149 \\ 
  &  &  &  &  &  &  &  &  & (0.0123) \\ 
 [-1.8ex] \\
  High:Eradication &  &  &  &  &  &  &  &  & $-$0.0632$^{**}$ \\ 
  &  &  &  &  &  &  &  &  & (0.0319) \\ 
 \hline \\[-1.8ex] 
Observations & 1,405 & 1,405 & 1,404 & 1,389 & 1,405 & 1,405 & 1,404 & 1,405 & 677 \\ 
R$^{2}$ & 0.1313 & 0.1528 & 0.2534 & 0.1141 & 0.2523 & 0.1028 & 0.2030 & 0.1057 & 0.1943 \\ 
Adjusted R$^{2}$ & 0.0939 & 0.1104 & 0.2212 & 0.0754 & 0.2200 & 0.0641 & 0.1686 & 0.0671 & 0.1201 \\ 
\hline 
\hline \\[-1.8ex] 
% \textit{Note:}  & \multicolumn{9}{r}{$^{*}$p$<$0.1; $^{**}$p$<$0.05; $^{***}$p$<$0.01} \\ 
\end{tabular} 
  \caption{\textbf{Full regression results used to derive all Figures in main text.} Columns (3)-(4) and (5)-(8) display the regression results in Figure 4B and 4C respectively, from bottom-most bars to top-most bars. Coefficient estimates for the fixed effects are not displayed. $^{*}$p$<$0.1; $^{**}$p$<$0.05; $^{***}$p$<$0.01.}
\end{table} 

\clearpage
% Bibliography
% \bibliographystyle{naturemag}
% \bibliography{displacement,migration,poppy,targeting, afg_qual}

\end{document}